\documentclass{CUP-JNL-EDS}

\usepackage{latexsym}
\usepackage{graphicx}
\usepackage{multicol,multirow}
\usepackage{amsmath,amssymb,amsfonts}
\usepackage{mathrsfs}
\usepackage{amsthm}
\usepackage{rotating}
\usepackage{appendix}
\usepackage{natbib}
\usepackage{ifpdf}
\usepackage[T1]{fontenc}
\usepackage{tikz}
\usepackage{comment}

\usetikzlibrary{decorations.pathreplacing, positioning, decorations.pathmorphing}
\usetikzlibrary{matrix}
\usepackage{tikz-3dplot}
\usetikzlibrary{3d}

\usetikzlibrary{tikzmark, positioning, fit, shapes.misc, arrows.meta,calc}
\usepackage[export]{adjustbox}
\usetikzlibrary{intersections}
\usepackage[skins]{tcolorbox}

\tikzstyle{mybox} = [draw=red, fill=orange!20, very thick,
rectangle, rounded corners, inner sep=10pt, inner ysep=20pt]
\tikzstyle{fancytitle} =[fill=orange!50, text=black]

\usepackage{times}
\usepackage{newtxmath}
\usepackage{textcomp}%
\usepackage{xcolor}%
\usepackage{hyperref}
\usepackage{lipsum}
\usepackage[linesnumbered,ruled]{algorithm2e}
\usepackage{subcaption}
\usepackage{comment}
\usepackage{tabularray}
\UseTblrLibrary{booktabs}
\definecolor{GMcolor}{rgb}{0.1,0.1,1}
\definecolor{PYcolor}{rgb}{0.5,0.1,1}
\definecolor{FBcolor}{rgb}{1,0.1,0.1} 
\definecolor{DLcolor}{rgb}{1,0.5,0.1}

\articletype{}
\jname{Submitted to Environmental Data Science}
\jdoi{}

\begin{document}

\begin{Frontmatter}

\title{Extreme heatwave sampling and prediction with analog Markov chain and comparisons with deep learning}

\author[1,2]{George Miloshevich }\orcid{0000-0002-8514-4315}
\author[1,3,4]{Dario Lucente}\email{name2@email.com}
\author[2]{Pascal Yiou}\orcid{0000-0001-8534-5355}
\author[1,5]{Freddy Bouchet}

\address*[1]{\orgdiv{ENSL}, \orgname{CNRS, Laboratoire de Physique}, \orgaddress{\city{Lyon},  \country{France}}}

\address[2]{\orgdiv{Laboratoire des Sciences du Climat et de l’Environnement, UMR 8212 CEA-CNRS-UVSQ}, \orgname{Université Paris-Saclay \& IPSL, CE Saclay Orme des Merisiers}, \orgaddress{\city{Gif-sur-Yvette},  \country{France}}}

\address*[3]{\orgdiv{Department of Physics}, \orgname{University of Rome Sapienza, P.le Aldo Moro 2, 00185}, \orgaddress{\city{Rome}, \country{Italy}}}

\address*[4]{\orgdiv{Institute for Complex Systems}, \orgname{CNR, P.le Aldo Moro 2, 00185}, \orgaddress{\city{Rome},\country{Italy}}}

\address*[5]{\orgdiv{LMD/IPSL}, \orgname{ ENS, Université PSL, École Polytechnique, Institut Polytechnique de Paris, CNRS} \orgaddress{\city{Paris},  \country{France}}}


\authormark{Miloshevich et al.}


\abstract{We present a data-driven emulator, a stochastic weather generator (SWG), suitable for estimating probabilities of prolonged heatwaves in France and Scandinavia. This emulator is based on the method of analogs of circulation to which we add temperature and soil moisture as predictor fields. We train the emulator on an intermediate complexity climate model run and show that it is capable of predicting conditional probabilities (forecasting) of heatwaves out of sample. Special attention is payed that this prediction is evaluated using a proper score appropriate for rare events. To accelerate the computation of analogs dimensionality reduction techniques are applied and the performance is evaluated. The probabilistic prediction achieved with SWG is compared with the one achieved with a Convolutional Neural Network (CNN). With the availability of hundreds of years of training data CNNs perform better at the task of probabilistic prediction. In addition, we show that the SWG emulator trained on 80 years of data is capable of estimating extreme return times of order of thousands of years for heatwaves longer than several days more precisely than the fit based on generalised extreme value distribution. Finally, the quality of its synthetic extreme teleconnection patterns obtained with SWG is studied. We showcase two examples of such synthetic teleconnection patterns for heatwaves in France and Scandinavia that compare favorably to the very long climate model control run.}

\policy{Estimating conditional probabilities and rate of returns of extreme heatwaves is important for climate change risk assessments. The most impactful heatwaves are also long-lasting. Numerical weather and climate models are expensive to run and may have biases, while observational datasets are too short to observe many of the potential extreme events and sample them accordingly. We present a relatively inexpensive yet powerful data-driven tool for sampling and estimating probabilities of extreme prolonged heatwaves and compare it with existing data-driven statistical approaches such as convolutional neural networks and extreme value theory estimations.
 }

\end{Frontmatter}

\section{Introduction}

It is expected that heatwaves will be among the most impactful effects of climate change~\citep{russo2015top,IPCC_2021_extremes} in the 21st century. 
While, generally, climate scientists have warned that global warming will increase the rate of heatwaves, in some regions such as Europe the trends have accelerated beyond what was expected~\citep{Christidis2020}. 
Extreme heatwaves such as the Western European heatwave in 2003~\citep{Herrera2010} and the Russian heatwave in 2010~\citep{barriopedro2011hot,Chen22}  had consequences on agriculture and posed health hazards. Such events have also affected other regions including Asia, e.g. the Korean Peninsula~\citep{Choi2021} in the summers of 2013, 2016 and 2018~\citep{ki20}. Thus, understanding the potential drivers, forecasting as well as performing long-term projections of heatwaves is necessary. These tasks, also relevant for climate change attribution studies~\citep{national_academies_of_sciences_engineering_and_medicine_attribution_2016}, are difficult since they require computation of small 
probabilities and thus massive ensemble simulations with expensive models 
or drawing conclusions from the relatively short observational record. 

The possibility of using atmospheric circulation analogs for predicting weather dates back to the works of \citep{bjerknes1921meteorology} and~\citep{lorenz69}. The idea is quite intuitive: one expects weather patterns to repeat given similar initial conditions. 
However, estimating the amount of data (a \emph{catalog} of analogs) actually necessary for short range predictions revealed superiority of physics-based Numerical Weather Prediction (NWP) models ~\citep{balaji,vandenDool07}. On the other hand, for predicting certain types of large-scale patterns analog forecasting may require less resources and can be competitive with or superior to NWP on time scales longer than Lyapunov time such as Subseasonal-to-Seasonal (S2S)  scales~\citep{van2003performance,cohen19} and especially long-range predictions. The examples include prediction of ENSO~\citep{ding2019diagnosing,wang2020extended} and Madden-Julian Oscillation~\citep{krouma23}.

Analog forecasting involves constructing potential trajectories  that the system could have explored from a given state. These methods, which assume a Markov property~\citep{yiou2014anawege} belong to the family of methods referred to as Stochastic Weather Generators (SWGs). 
SWGs can be used as climate model emulators due to their ability to generate long sequences that can infer statistical properties of spatio-temporal dynamics and correlation structures of the system without running expensive General Circulation Models (GCMs)~\citep{ailliot2015stochastic}. Other applications include down-scaling~\citep{rajagopalan1999k,wilks1992adapting} and data assimilation~\citep{lguensat2017}.  

In the last decade there has been a rapid advancement of other types of data-driven forecasting/emulators in earth system models based on deep learning~\citep{reichstein19}. For example, in a seminal work~\citep{Ham} Convolutional Neural Network (CNN) has achieved a positive skill for ENSO prediction compared to NWP. These success stories are sometimes accompanied by similar skills displayed by the analog method~\citep{ding2019diagnosing,wang2020extended}.  Recently, deep learning has been used to predict extreme heatwaves~\citep{Chattopadhyay19,jacques-dumas22,lopez2022global} targeting categorical scores related to hit rates\footnote{An example of categorical score is Matthews Correlation Coefficient (MCC) which was designed as a measure of categorical classification and is based on a combination of true/false positive/negatives appropriate for imbalanced datasets.}. In order to provide probabilistic prediction the following studies were performed using random forests~\citep{straaten22} and neural networks~\citep{miloshevich22}. Probabilistic forecasting in the context of uncertainty quantification plays an important role in the current development of machine learning driven techniques applied to, for instance, post-processing of weather forecasts~\citep{shulz22,gronquist_deep_2021}. 

Geopotential height anomalies and soil moisture were chosen as the inputs in~\citep{miloshevich22} based on physical understanding of the precursors to heatwaves.
Persistent positive geopotential height anomalies 
are known drivers, since 
they favor clear skies and produce subsidence.  It has been argued that soil moisture has memory of previous land-atmospheric conditions~\citep{koster2001soil,seneviratne2006soil}. In fact, soil moisture was identified~\citet{Seneviratne12} among the important drivers for European heatwaves. In contrast, in Northern European regions soil moisture may play a lesser role in preconditioning~\citep{felsche23}. This association between soil moisture deficits and heatwave occurrence has been indicated also elsewhere~\citep{shukla1982influence,rowntree1983simulation,dandrea06,vautard2007,seneviratne2010investigating,fischer2007soil,lorenz2010persistence,Stefanon_2012,hirschi2011observational,schubert2014northern,zhou2019land,benson2021characterizing,zeppetello2022physics,vargas2020projected,miralles14,miralles19} contributing to the understanding that in dry regimes heatwaves could be amplified due to impacts of evapotranspiration. 
In contrast, the SWG, designed for heatwaves, did not take this crucial input and thus was not able to reproduce the corresponding feedback~\citep{jezequel2018role}. However, it is important to better understand the contributions of such drivers to improve projected climate change impacts on heatwave probabilities~\citep{Berg15,field2012managing,horton2016review}. 
{For instance, using circulation analogs in a pre-industrial simulation,~\citep{horowitz2022circulation} showed that circulation patterns explain around 80\% of the temperature anomalies in the United States, while negative soil moisture anomalies added a significant positive contribution, especially mid-continent.}

Recently, the analog method was successfully adapted for predicting chaotic transitions in a low-dimensional system~\citep{Lucente2022committor}. In fact, the analog method is generally expected to perform better when the number of relevant degrees of freedom is not too high. This is natural given that it belongs to the family of $k$-nearest neighbors algorithms, which suffer from the curse of dimensionality~\citep{beyer1999nearest}.  Consequently, for our problem of heatwave prediction we will consider linear and nonlinear dimensionality reduction techniques and evaluate the performance of SWG in real vs latent space. This approach is partially motivated by the emergence of generative modelling for climate and weather applications, e.g. studies combining deep learning architectures with Extreme Value Theory (EVT) for generating extremes~\citep{Bhatia21,boulaguiem22} and realistic climate situations \citep{besombes2021producing}. 

Beyond finite-horizon probabilistic prediction, data-driven methods can be used to create emulators capable of extracting risks for ``black-swan`` events, events that are so far removed from the climatology that they have not been observed yet but their impact may be devastating. Such risk assessments are often carried out using EVT based on reanalysis, with long runs of high fidelity models (methods such as ensemble boosting~\citep{gessner21}) or rare event algorithms~\citep{Ragone18,Ragone21}. More recently, other statistical approaches such as Markov state models~\citep{finkel23a} have been used to estimate the rates of extreme sudden stratospheric warming events based on ensemble hindcasts of European Center of Medium Range Weather Forecasting (ECMWF).  

The first goal of this study is to compare the performance of two data-driven probabilistic forecasting approaches for extreme heatwaves in two European regions: France and Scandinavia. We aim to explore the use of Stochastic Weather Generator (SWG), that relies on the method of analog Markov chain, optimize it for our task and compare or combine it with more modern deep learning approaches such as CNN.  We will investigate ways to accelerate the computation of analogs using dimensionality reduction techniques, such as training a Variational Autoencoder (VAE) to project the state of the system to a small-dimensional latent space. Our second goal will consist of generating synthetic time series with the help of the SWG trained on a short model run and comparing statistics of under-resolved extreme events (in this short run) to a long control run. We will work with PlaSim data, an intermediate complexity General Circulation Model (GCM). The choice of the right metrics is important since many of the current data-driven forecasts are trained based on Mean Square Error (MSE) which is not suited for evaluation of the extremes. However, extremes actually represent the most immediate societal risks~\citep{watson_machine_2022} and are a subject of this manuscript. The choice of GCM is motivated by the ability of PlaSim to generate inexpensive long runs and the recent study of~\cite{miloshevich2023robust} who showed that the composite statistics of the large scale 500 hPa geopotential height fields, conditioned on heatwaves, revealed similar patterns for PlaSim as for higher fidelity models such as CESM and the ERA5 reanalysis.

The paper is organised as follows: Section 2.1 describes the details of our dataset generated by Plasim simulation, Section 2.2 defines notation and heatwave events. Section 2.3 delineates the goal of the probabilistic inference, the scoring function which determines the goodness of the predictions, the training and validation protocol. Section 2.4 reviews the CNN introduced in~\citet{miloshevich22}, Section 2.5 introduces analog Markov chain, i.e. SWG and the corresponding steps involving coarse-graining, definition of metric, how to make probabilistic prediction with SWG and how to construct return time plots. Section 3 spells out the results consisting of Section 3.1 that covers probabilistic forecasting and Section 3.2 which discusses extending return time plots and teleconnections. Finally Section 4 concludes the findings of this study.

\section{Methods and data}

\subsection{PlaSim model data}

\begin{figure}
    \centering
    \includegraphics[width=0.75\linewidth]{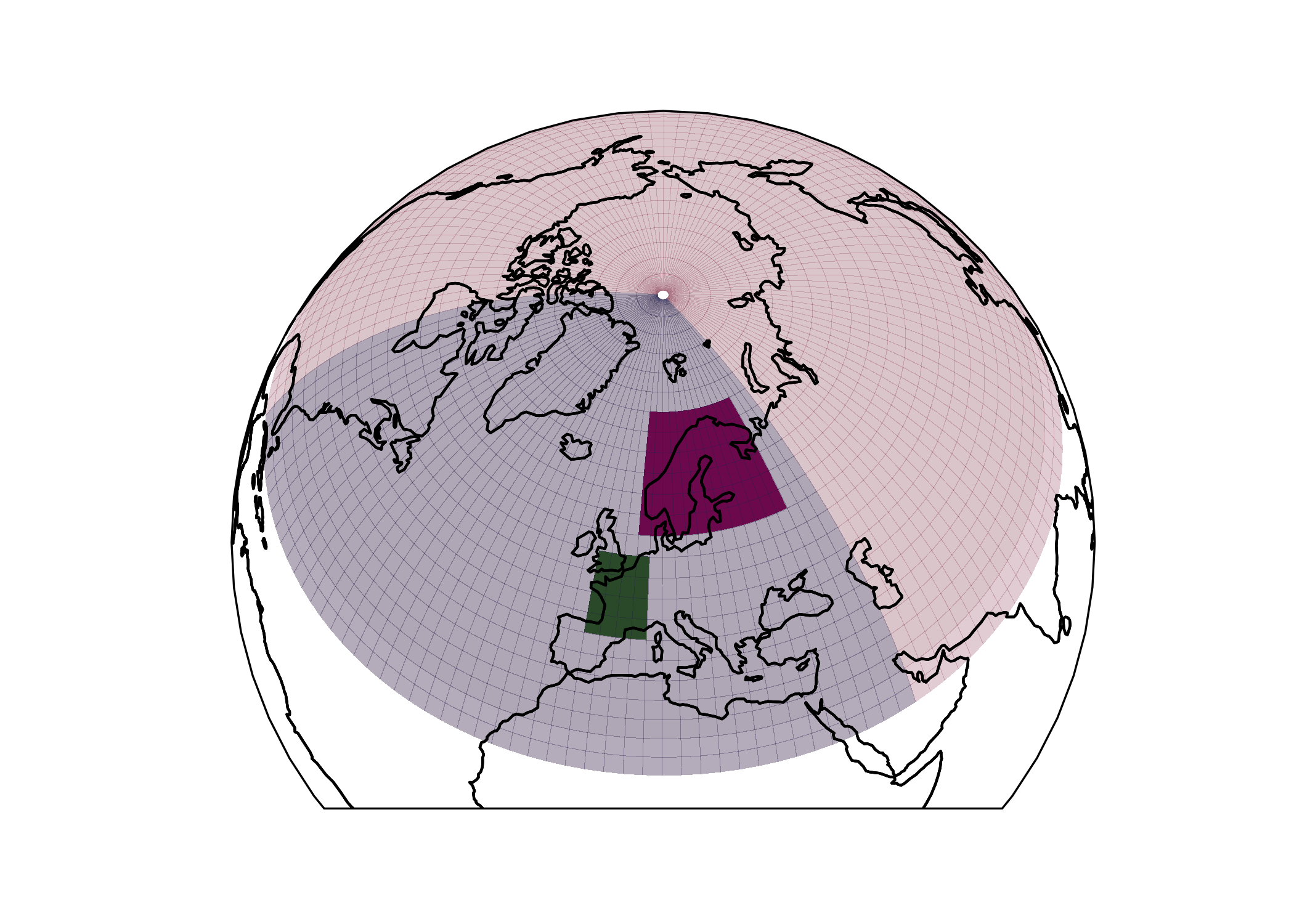}
    \caption{The map shows relevant areas. Blue:  North Atlantic -- Europe (NAE). The dimensions of this region are 22 by 48. Red+Blue: North Hemisphere (above 30 N). The dimensions of this region are 24 by 128. Green: France; Purple: Scandinavia.}
    \label{fig:Plasim_areas}
\end{figure}

The data have been generated from 80 batches\footnote{In this context ``batch'' refers to independent runs, i.e. 80 independent initial conditions. In contrast, it does not refer to the procedure of splitting a dataset into portions that is performed during the gradient descent when training neural network.} that are independent 100 year long stationary simulations of PlaSim~\citep{fraedrich2005planet,fraedrich_1998}. PlaSim is an intermediate complexity General Circulation Climate Model (GCM), suitable for methodological developments, like the one performed here. It can be run to generate long simulations at a lower computational cost than the new generation of CMIP6 ensemble. 


PlaSim consists of an atmospheric component coupled to ice, ocean and land modules. The atmospheric component solves the primitive equations of fluid dynamics adapted to the geometry of the Earth in the spectral space. Unresolved processes such as boundary heat fluxes, convection, clouds, etc are parameterized. The interactions between the soil and the atmosphere are crucial for this study. They are governed by the bucket model of~\citet{manabe69}, which controls the soil water content by replenishing it from precipitation and snow melt and depleting by the surface evaporation. The soil moisture capacity is prescribed based on geographical distribution.

The model was run with conditions corresponding to the Earth climate in 1990s with fixed greenhouse gas concentrations and boundary conditions which include incoming solar radiation, sea surface temperatures and sea ice cover that are cyclically repeated each year. The parameters are chosen to reproduce the climate of the 1990s. We also include the daily cycle in this work (like in~\citet{miloshevich22}). The model is run at T42 resolution, which corresponds to cells that are $2.8$ by $2.8$ degrees and results in $64$ by $128$ resolution over the whole globe. The vertical resolution is 10 layers. The fields are sampled every 3 hours and daily averages are taken.

\begin{table}
  \centering
  \begin{tblr}{
      colspec={lll},
      row{1}={font=\bfseries},
      column{1}={font=\itshape},
      row{even}={bg=gray!20},
    }
      abbreviation  & training years & validation years \\
    \toprule
    $D100$ & 80 & 20 \\
    $D500$ & 400 & 100 \\
    $D8000$ & 7200 & 800 \\
    \bottomrule
  \end{tblr}
  \caption{Break-up of the input data into training and validation for different subsets of the full D8000 dataset. {In D100 and D500 we follow 5-fold cross-validation, while in D8000 exceptionally 10-fold cross-validation was employed. The latter implies that we slide the test window 10 times so that it
explores the full dataset, while the training set is always the complement.}}
  \label{tab:datasets}
\end{table}

In this paper we work with different subsets of the full 8000 year long dataset defined in  Table~\ref{tab:datasets}. 


\subsection{Physical fields and geographical domains}\label{sec:inputs_domains}

The inputs to our various data pipelines will consist of fields $\mathcal{F}(\mathbf{r},t)$ such as
\begin{itemize}
    \item $\mathcal{T}$: 2-meter temperature 
    \item $\mathcal{Z}$: 500 hPa geopotential height 
    \item $\mathcal{S}$: soil moisture 
\end{itemize}
at time $t$,  and discrete points $\mathbf{r}\in \mathbb{D}$, where $\mathbb{D}$ represents domain of interest 
. We will consider three domains: 
\begin{itemize}
    \item $\mathbb{D}_{NAE}$: North Atlantic, Europe (24 by 48 cells)
    \item $\mathbb{D}_{France}$: the area of France masked over the land
    \item $\mathbb{D}_{Scandinavia}$: the area of Scandinavia masked over the land
\end{itemize}
The areas of France and Scandinavia that will correspond to the geographical areas of heatwaves are defined in Figure~\ref{fig:Plasim_areas}. 
Data-driven methods that will be introduced below will accept typically some stacked version of the fields which will be represented by the letter $\mathcal{X}$

\begin{equation}\label{eq:input_X}
    \mathcal{X} = \left({\mathcal{Z}}(\mathbf{r},t), {\mathcal{T}}(\mathbf{r},t), {\mathcal{S}}(\mathbf{r},t) \right),
\end{equation}
where each component corresponds to a specific field.

\subsection{Definition of heatwave events}
As in~\citep{miloshevich22} we are concerned with the prediction of {$T=15$} day-long heatwaves. Thus, our events of interest 
are related to the time and space integrated anomaly of 2-meter temperature ($\mathcal{T}$), where we use symbol $\mathbb{E}$ to imply average over the years conditioned to the calendar day:

\begin{equation}\label{eq:timeaveraged}
	A(t)
	=\frac{1}{T}\int_{t}^{t+T}\frac{1}{\mathbb{\left|D\right|}}\int_{\mathbb{D}}\left(\mathcal{T}-\mathbb{E}\left(\mathcal{T}\right)\right)(\boldsymbol{r},t^\prime)\,\mathrm{d}\boldsymbol{r}\,\mathrm{d}t^\prime,
\end{equation}
which depending on the threshold $A(t) \ge \alpha$ defines a class of heatwaves. Symbol $\mathbb{D}$ corresponds to the domain of interest described in Section~\ref{sec:inputs_domains}. $\mathbb{\left|D\right|}$ is the surface {area} of $\mathbb{D}$ {and equation~\eqref{eq:timeaveraged} contains an area weighted average}. $T$  is the chosen duration (in days) of the heatwaves. For probabilistic prediction we set $T=15$ days, while for return times we relax this requirement to study a family of return time plots. 

The paper will involve two slightly different definitions of heatwaves based on equation~\eqref{eq:timeaveraged} depending on the task that will be performed. In the next Section~\ref{sec:committor} we will introduce probabilistic prediction, for which heatwaves will be defined as exceeding a certain fixed threshold $a$ set fixed for all heatwaves, which allows many heatwaves per summer. Concurrently, in Section~\ref{sec:returns_methods} we will work with a block-maximum definition, which permits only single heatwave per summer, so each year gets unique value $a$. Moreover, a range of $T$ values will be considered.

\subsection{Probabilistic prediction and validation
}\label{sec:prob_pred_val}

\subsubsection{Committor Function}\label{sec:committor}

Our goal is a probabilistic forecast, a \emph{committor function}~\citep{Lucente2022committor} in the nomenclature of rare event simulations. 
Based on equation~\eqref{eq:coarsetimeaveraged} we  define the label $y=y(t)$ for each state as a function of the  corresponding time $t$ as $y(t) \in \{0,1\}$, so that $y=1$ iff $A>a$, where $a$ is $95$-th percentile of $A$. Our objective is to find the probability $p$ of $y=1$ at time $t$ given the  state $\mathcal{X}$ which we observe at an earlier time $t-\tau$ (see Section~\ref{sec:state_space_metric}):
\begin{equation}\label{eq:committor}
    p = \Pr(y(t)=1|\mathcal{X}(t-\tau),\tau).
\end{equation}
Parameter $\tau$ is referred to as lead time or sometimes lag time. We stress that for non-zero $\tau$ it is the initial observed state $\mathcal{X}$ that is shifted in time, rather than the labels $y$, thus the events of interest (both $y=1$ or $y=0$) are kept the same, allowing controlled comparisons among different values of $\tau$. Moreover, the season of interest for which we intend to compute the committor functions and compare them to the ground truth should always involve the same days: from June 1 to August 16, because the last potential heatwave may last $T = 15$ days and, this way, end in August 30 (the last day of summer in PlaSim). 


\subsubsection{Normalized Logarithmic Skill Score}\label{sec:NLS}
The quality of the prediction will be measured with a \emph{Normalized Logarithmic Score} (NLS) ~\citep{miloshevich22} due its attractive properties for rare events~\citep{benedetti10} and the appropriateness for probabilistic predictions~\citep{wilks19}. 
{In the following we will briefly introduce the NLS. Since the dataset is composed of $N$ states $\left\{\mathcal{X}_n\right\}_{1\ge n \ge N}$, each of them labeled by $y_n\in\{0,1\}$, it can be equivalently represented by $N$ independent pairs $\left(\mathcal{X}_n,y_n\right)$. The aim of the prediction task is to provide an estimate $\hat{p}_y$ of the unknown probability $p_{\bar{y}}(X)=\mathbb{P}(y=\bar{y}|\mathcal{X}=X)$. It has been argued~\citep{miloshevich22,benedetti10} that among the possible scores to assess the quality of a probabilistic prediction, the most suitable is the logarithmic score \begin{equation}
	S_N(\hat{p}_y) =- \frac{1}{N}\sum_{n =1}^N \log(\hat{p}_{y_n}(\mathcal{X}_n))\,.
\end{equation}
It should be noted that the logarithmic score coincides with the empirical cross entropy widely used in machine learning. The score $S_N$ is positive and negatively oriented (the smaller the better). The value of $S_N$ is not indicative in itself, but is useful for comparing the quality of two different predictions. Thus, the NLS is defined as an affine transformation of the logarithmic score, which allows us to compare the data-driven forecast with the climatological one. To be more precise, let $\bar{p}$ be the climatological frequency of the extremes ($\bar{p}= \mathbb{P}(y=1)$) and $S_N^{(c)}$ the logarithmic score for this prediction, i.e.
\begin{equation}
	S_N^{(c)}(\bar{p}) =- \frac{1}{N}\sum_{n=1}^N \left[ \delta_{y_{n},1}\log(\bar{p})+\delta_{y_{n},0}\log(1-\bar{p})\right]=-\bar{p}\log(\bar{p})-(1-\bar{p})\log(1-\bar{p})\,.
\end{equation}
Then, the NLS score is defined as
\begin{equation}\label{eq:NLS}
	\text{NLS}(\hat{p}_y) = \frac{S_N^{(c)}-S_N(\hat{p}_y)}{S_N^{(c)}(\bar{p})}.
\end{equation}
Thus, NLS is positively oriented (the larger the better) and bounded above by $1$.
}
Since we identify $y=1$ with 95th percentile of $A(t)$ this implies that $\bar{p} = 0.05$. In other words, we always compare the results to the baseline\footnote{In this case baseline would be always guessing the heatwave will occur with 5 percent probability independent of X}, which would result in NLS equal to zero. Any predictions with smaller NLS are completely useless from the probabilistic perspective. Finally, we stress that other scores devised for rare events, such as MCC~\citep{MATTHEWS1975}, most notably used in the study~\citep{jacques-dumas22} and others, are useful for categorical prediction but do not necessarily translate into high NLS scores when training neural networks especially if early stopping\footnote{early stopping refers to the callback which stops adapting the weights of the neural network once an objective criterion has been reached which is often measured by monitoring the loss function. } is used, since the optimal epoch may vary depending on the chosen score. 


\subsubsection{Training/Validation protocol for committor function}\label{sec:training_validation}

We split the dataset into training (TS) and validation (VS). The ML algorithm of choice is trained on TS and various hyper-parameters are optimized on VS depending on the target metric (see the Section~\ref{sec:NLS}). Different methods can be compared on VS, which will be performed in this paper. The splits will be based on the Table~\ref{tab:datasets} {and correspond to the selection of the first number of years, e.g. D100 implies choosing the first 100 years of the simulation. }

To have a better idea about the spread of the skill (equation~(\ref{eq:NLS})) we will rely on 5-fold cross-validation. This means that the TS/VS split is performed 5 times, while sliding the VS window consequently through the full data interval, chosen for the particular study (D100, D500 etc). Each TS/VS split we refer to as ``fold'', i.e. there are 5 folds {(numbered 0-4)}, for instance for D100, fold number 3, VS starts in year 60 and ends in year {79}, while TS is as usual the complement of that set. {Notice that years are also numbered from 0}. In order to ensure that each VS has the same number of extreme events we perform custom stratification~\citep{miloshevich22}. This means that we shuffle the years of the simulation between different folds so that each interval receives the same (or almost the same) number of extreme events. The procedure does not modify the mean benchmarks but tends to decrease the error bars, which strongly depend on the number of extreme events within the VS.


\subsection{Convolutional Neural Network}\label{sec:CNN}

\begin{figure}
	\includegraphics[width=1.1\textwidth]{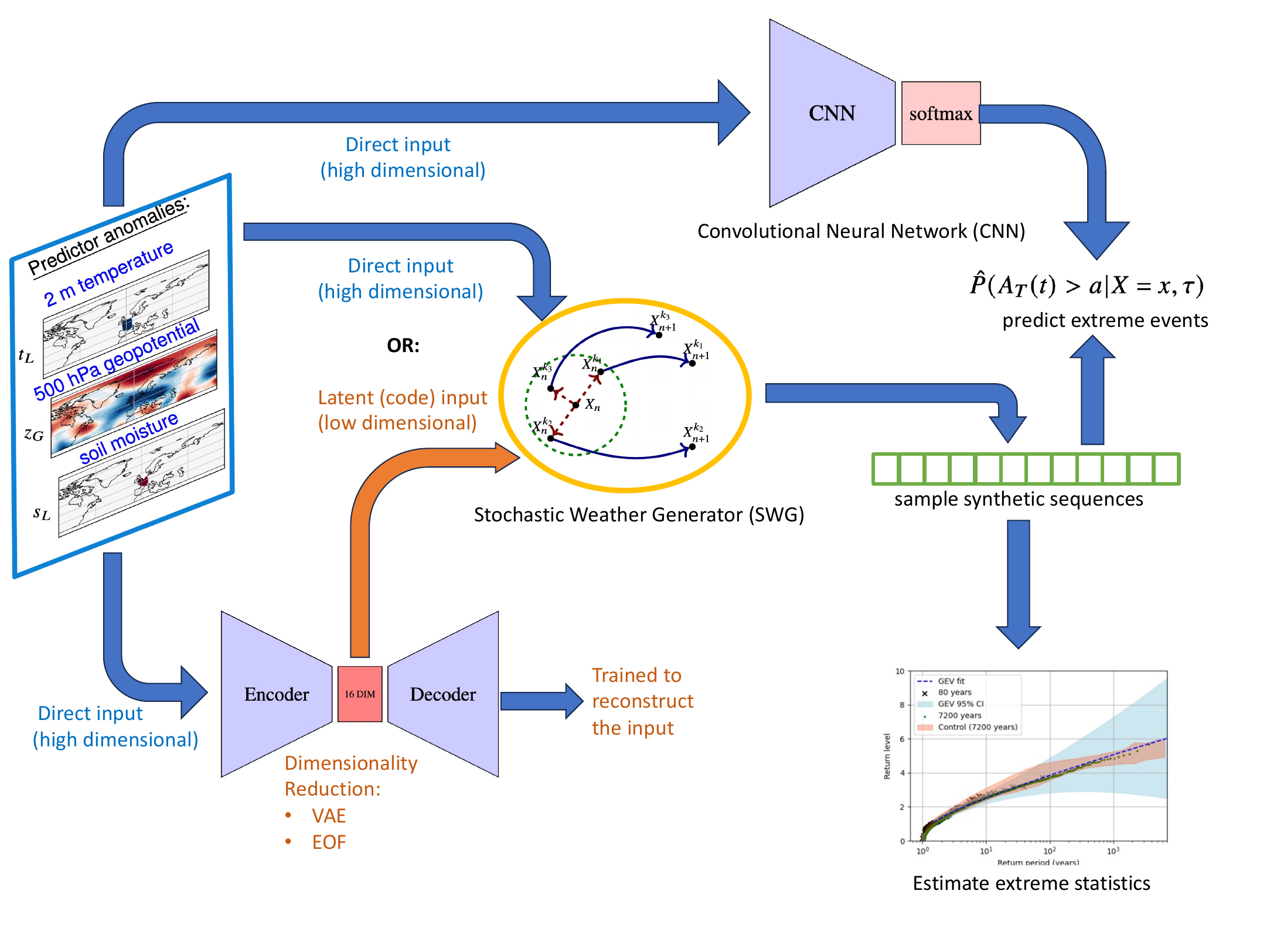}
	\caption{Schematics of different methodologies used for probabilistic forecasting {and estimates of extreme statistics}. On the left we show the three input fields which are labeled directly on the plot. On the right the target of the inference is displayed {(for instance probabilistic prediction as described in ~\ref{sec:SWGcommittor})}. On top we have the direct CNN approach. In the middle we show the analog method (SWG), which is the main topic of this work and which is compared to CNN for this task. At at the bottom the option is presented to perform dimensionality reduction of the input fields and pass them as the input on which SWG is ``trained''. {Subsequently, SWG can be used to generate conditional or unconditional synthetic series (green boxes). This synthetic data can be used for make probabilistic prediction or estimating tails of distribution, such as return time plots} }
	\label{fig:architecture}
\end{figure}

The Convolutional Neural Network (CNN) we take for this study has been proposed by~\citet{miloshevich22}. Since the precise inputs are slightly different in this paper (as well as a whole new region of Scandinavia that is considered in this paper for the first time) the training had to be repeated, but the hyperparameters were left unchanged. 

In this manuscript the CNN accepts an input that is 24 by 48 grid-points over three fields (see Eq.~\eqref{eq:input_X}). 
We perform masking~\citep{miloshevich22} of the temperature and soil moisture, i.e. values outside of the respective heatwave area (France or Scandinavia depending on which heatwaves we study, see Figure~\ref{fig:Plasim_areas}) are set to zero to avoid some spurious correlations. In this paper we also study the heatwaves occurring in Scandinavia, and for consistency perform masking over the corresponding area in that case. The inputs are then normalized so that each grid-cell of each field has mean 0 and standard deviation 1. 


The input is fed through three convolutional and two max-pooling layers; the intermediate output is flattened and passed through a dense layer, which is connected to the final output. The final output consists of two neurons and a soft-max activation, which enforces the outputs within $(0,1)$ interval consistent with the domain of probabilities. 
The architecture is trained with Adam optimizer to minimize cross-entropy as is appropriate for classification task and early stopping is applied which stops the training when the average foldwise NLS reaches the maximum. 

\subsection{Analog Markov chain}\label{sec:analog_method}


\subsubsection{Stochastic Weather Generator algorithm}\label{sec:SWG_algorithm}

\begin{figure}
    \centering
    	\begin{tikzpicture}[scale=0.8]
	      \begin{scope}[shift={(-1,-5)}]
		    	\draw[dashed, color=green!50!black, ultra thick] (0,0) circle (1.5cm);
		    	\node[circle, draw, fill=black, inner sep=1.5pt, label=right:$X_n$] (Xn) at (0,0) {};
		    	\node[circle, draw, fill=black, inner sep=1.5pt, label={[xshift=-4mm, yshift=-3mm]$X_n^{k_1}$}] (Xnk1) at (0.75,1) {};
		    	\node[circle, draw, fill=black, inner sep=1.5pt, label={}] (Xn1k1) at (4,0) {};
		    	
		    	\node[above left=2cm of Xn, inner sep=0, align=center,left=2cm,  above=1.5cm, xshift=0cm] {
		    		Validation Set
		    	};
		    	
		    	\draw[->, color=red!50!black, ultra thick, dashed] (Xn) -- (Xnk1);

		    	\draw[->, color=blue!50!black, ultra thick] (Xnk1) to[out=20,in=160] (Xn1k1);

	    \end{scope}
    		 \begin{scope}[shift={(3,-5)}]
    			\draw[dashed, color=green!50!black, ultra thick] (0,0) circle (1.5cm);
    			\node[circle, draw, fill=black, inner sep=1.5pt, label=right:$X_{n+1}$] (Xn) at (0,0) {};
    			
    			\node[circle, draw, fill=black, inner sep=1.5pt, label={[xshift=-1mm, yshift=1.5mm]$X_{n+1}^{k_2}$}] (Xnk2) at (-0.75,-1) {};
    			\node[circle, draw, fill=black, inner sep=1.5pt, label={}] (Xn1k2) at (4,0) {};
    			
    			\node[above left=2cm of Xn, inner sep=0, align=center,left=2cm, above=1.5cm, xshift=0cm] {
    				Training Set
    			};
    			
    			\draw[->, color=red!50!black, ultra thick, dashed] (Xn) -- (Xnk2);
    			
    			\draw[->, color=blue!50!black, ultra thick] (Xnk2) to[out=-20,in=-160] (Xn1k2);
    			
    		\end{scope}
    		 \begin{scope}[shift={(7,-5)}]
    			\draw[dashed, color=green!50!black, ultra thick] (0,0) circle (1.5cm);
    			\node[circle, draw, fill=black, inner sep=1.5pt, label=right:$X_{n+2}$] (Xn) at (0,0) {};
    			
    			\node[circle, draw, fill=black, inner sep=1.5pt, label={[xshift=-1.5mm, yshift=0mm]$X_{n+2}^{k_3}$}] (Xnk3) at (-0.75,0.5) {};

    			\node[circle, draw, fill=black, inner sep=1.5pt, label={}] (Xn1k3) at (4,0) {};
    			\node at (4.75,0) {$X_{n+3}$};
    			\node[above left=2cm of Xn, inner sep=0, align=center,left=2cm, above=1.5cm, xshift=0cm] {
    				Training Set
    			};

    			\draw[->, color=red!50!black, ultra thick, dashed] (Xn) -- (Xnk3);
    			
    			\draw[->, color=blue!50!black, ultra thick, bend left=50] (Xnk3) to (Xn1k3);

    		\end{scope}
	\end{tikzpicture}
    \caption{The schematics of the flow of analogs. When applying the algorithm for estimation of committor (equation~\eqref{eq:committor}) the first step consists of starting from the state in the Validation Set (VS), finding the analog in the Training Set (TS) and applying the time evolution operator.  All subsequent transition occur within the TS.}
    \label{fig:analogs}
\end{figure}
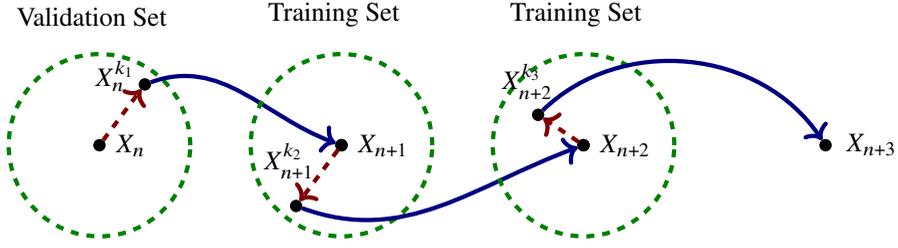

The principle of the Stochastic Weather Generator (SWG) as described in~\citep{yiou2014anawege} is based on the analog Markov chain method. The idea is to use the existing observations or model output that catalog the dynamical evolution of the system, \emph{the dynamical history}, and generate synthetic series. 

\begin{algorithm}
    \SetKwInOut{Input}{Input}
    \SetKwInOut{Output}{Output}
    \underline{Define variables} $(K,N_t,\mathcal{M}^{N_t\times K})$\;
    \Input{$K \gets$ the number of nearest neighbors retained}
    \Input{$N_t \gets$ the number of states in the training set}
    \Input{Compute the matrix of \underline{nearest neighbors} $\mathcal{M}^{N_t\times K}$}
    \Input{$\mathcal{T}=\mathcal{T}(n)\gets$ temperature sequence corresponding to historical states $n\in 1\dots N_t$}
    \Input{$s\gets$ The initial state of the synthetic trajectory}
    \Output{synthetic time series}
    append $\mathcal{T}(s)$ to the synthetic temperature series
    
    \While{synthetic trajectory continues}
      {
        $k\gets \mathbb{U}(K)$ a uniform random number draw, i.e. $0<k<K$ \label{alg:uniform_draw}

        $s\gets \mathcal{M}^{N_t\times K}_{s,k}$ a random analog of the current state

        $s\gets s + \tau_m$ a state is advanced according to its historical dynamical evolution. 
        
        append $s$ to the synthetic trajectory sequence\;
        append $\mathcal{T}(s)$ to the synthetic temperature series
      }
      {
        return synthetic temperature series
      }
    \caption{The algorithm generates a single synthetic trajectory by using the analogs of training set (see Table~\ref{tab:datasets}). Sets of $K$ nearest analogues of each sample $X_n$ during dynamical evolution are computed based on the metric given in the equation~\eqref{eq:metric}. The states whose analogs we compute are constrained by the condition that their calendar days must start on May 15 and end on August 30 (there are 30 days per month in PlaSim). This results in $N_t = 105\cdot N_{\text{years}}$. There is additional condition on the upper bound of the calendar day of the analogs of these days which is August 27 preventing SWG from accessing days outside of the summer season, since the time step is $\tau_m=3$ days. The analogs are computed using parallelized kDTree search implemented in scipy packages of python once per parameter $\alpha_0$ (see equation~\eqref{eq:metric} and discussion in Section~\ref{sec:state_space_metric}) and each fold. }
    \label{alg:SWG}
\end{algorithm}
The method consists of constructing a \emph{catalog of analogs}, the states which have similar circulation patterns and other thermodynamic characteristics (in our case temperature and soil moisture). Each state in the catalog comes from the realization of the dynamics (in priciple it could come not only from simulations but also reanalysis). To construct synthetic time series one starts from any of these states and draws randomly (with uniform probability) one analog from a pre-defined set of nearest neighbors $n$. The similarity between the states is assessed via Euclidean metric in the feature space~\citep{karlsson1987nearest,Davis02,yiou2014anawege}, as defined below (Section~\ref{sec:state_space_metric}). The following step is to look-up in the dynamical history, i.e. how that analog evolved dynamically after time $\tau_m$ and use the corresponding state as the next member of the synthetic time series (see Figure~\ref{fig:analogs}). Next, the process is repeated. The details of how the algorithm is implemented will become clear in the following sections but they are schematically represented on Figure~\ref{fig:analogs} and encapsulated procedurally in the Algorithm~\ref{alg:SWG}. {The procedure is quite general although it is still possible to modify it in many aspects. For instance, the analogs are not always drawn from the uniform probability~\citep{lall1996nearest,buishand2001multisite,rajagopalan1999k,yiou2014anawege}, the catalog of analogs is often constrained by a \emph{moving window} or the analogs from the same year may be excluded from possible candidates~\citep{yiou2014anawege}. The common rationale behind our choices is to prioritize methods that demand less resources and use as much information as possible. }

\subsubsection{Time step and coarse-graining}\label{sec:coarse}

The first question that arises is what is the appropriate choice for the time-step $\tau_m$ of SWG (See schematics on Figure~\ref{fig:analogs}). It is natural to use synoptic time scale of cyclones on which the dynamics partially decorrelates, which corresponds to $\tau_m = 3$ days
as demonstrated by inspecting autocovarience~\cite{miloshevich2023robust}. This means that we are constructing a Markov chain where we are selecting a time step consistent with ignoring synoptic time-scale correlations.   
{One may expect that the results do not depend too much on $\tau_m$ as long as it does not exceed the total correlation time (see Appendix~\ref{sec:daily} for further details). }
The coarse graining will be applied as follows,
$\tilde{\mathcal{F}}(\mathbf{r},t_0)$ will be
\begin{equation}\label{eq:coarse}
\tilde{\mathcal{F}}(\mathbf{r},t)=\frac{1}{\tau_c}\int_{t}^{t+\tau_c}{{\rm d}t_0\, \mathcal{F}(\mathbf{r},t_0)}. 
\end{equation}
The tilde will be suppressed without the loss of generality since the coarse-graining will be applied throughout, except in the input to the CNN. For simplicity, we perform coarse-graining in accordance with this time step $\tau_c = \tau_m$. 

Note that we will also work with scalars obtained as area integrals $\mathcal{F}_{\mathbb{D}}$ over the area of France $\mathbb{D}=F$ and Scandinavia $\mathbb{D}=S$ defined as
\begin{equation}\label{areaintegral}
    \langle\mathcal{F}\rangle_{\mathbb{D}}(t)=\frac{1}{\mathbb{D}}\int_{\mathbb{D}}{{\rm d}\mathbf{r}\, \mathcal{F}_{{\mathbb{D}}}(\mathbf{r},t)}.
\end{equation}
This allows us to re-write the definition of heatwaves (equation~\eqref{eq:timeaveraged}) in a new, more concise notation
\begin{equation}\label{eq:timeaveraged_notation}
	A(t)
	=
	\frac{1}{T}\int_{t}^{t+T} \mathrm{d}t^\prime \,\langle\mathcal{T} - \mathbb{E}(\mathcal{T})\rangle_{\mathbb{D}}(t^\prime) .
\end{equation}

If we take coarse-graining into account and set $T=n\tau_c=15$ days we can express equation~\eqref{eq:timeaveraged} as discretization
\begin{equation}\label{eq:coarsetimeaveraged}
A(t_0)=\frac{1}{n}\sum_{i=1}^n\langle\mathcal{T} - \mathbb{E}(\mathcal{T})\rangle_{\mathbb{D}}(t_0+(i-1)\tau_c).
\end{equation}

\subsubsection{State space and metric}\label{sec:state_space_metric}
Next, we address how to combine fields of different nature (unit) in SWG. The fields can be stacked in various ways, for instance tuples can be constructed:
\begin{equation}\label{eq:simple_markov}
    \mathcal{X} = \left(\overline{\mathcal{Z}}(\mathbf{r},t), \langle\overline{\mathcal{T}}\rangle_{\mathbb{D}}(t), \langle\overline{\mathcal{S}}\rangle_{\mathbb{D}}(t) \right).
\end{equation}
This approach adapted for SWG consisting of integrals over $\mathbb{D}$ is to be contrasted with the input received by CNN, whereby $\mathcal{S}$ and $\mathcal{T}$ are masked as described in Section~\ref{sec:CNN} outside areas corresponding to the heatwaves, while SWG receives only integrated temperature and soil moisture (over the same heatwave areas). 

In the case of SWG each field will be normalized to global field-wise standard deviation\footnote{The case of dimensionality reduction is addressed in the Appendix~\ref{sec:vae_methods}, where normalization is not necessary} (with the global field-wise mean subtracted) as opposed to cell-wise like in the case of CNN
\begin{equation}
    \overline{\mathcal{F}}(\mathbf{r},t) := \dfrac{{\mathcal{F}}(\mathbf{r},t)-\mathbb{E}[\langle{\mathcal{F}}\rangle_{\mathbb{D}}(t)]}{\sqrt{\left\langle Var[{\mathcal{F}}]\right\rangle_{\mathbb{D}}(t)}},
\end{equation}
The overbar will be omitted in what follows and it will be implied instead. 

The Euclidean distance between two points $\mathcal{X}_{1}$ and $\mathcal{X}_{2}$ is defined as
\begin{equation}\label{eq:metric}
    d(\mathcal{X}_{1},\mathcal{X}_{2}) := \sqrt{\int d\mathbf{r} \, \sum_i \frac{\alpha_i}{dim(\mathcal{X}^i)}\left(\mathcal{X}^i_1 - \mathcal{X}^i_2 \right)^2 },
\end{equation}
where $dim(\mathcal{X}^i)$ counts the number of grid points concerned (1 for scalars such as $\mathcal{S}$ and $\mathcal{T}$ and $1152$ for $\mathcal{Z}$). When $\alpha_1 = \alpha_2 = 0$ we get definition consistent with~\citep{yiou2014anawege}, except that we assign uniform probabilities for a set of nearest analogs.

Of a particular interest is the fact that geopotential contains global information on a shorter time-scale, while soil moisture contains local information on a longer time-scale \citep{miloshevich22}. In the zeroth order approximation one expects that for the efficient algorithm the fields would need to be further rescaled by their dimensionality as is done in equation~\eqref{eq:metric}. Thus, we choose parameters $\alpha_i = 1$ as a first guess. To optimise the performance of SWG for the conditional prediction problem (equation~\eqref{eq:committor}) we perform grid search: for each $\tau$ we compute committor $p$ and measure its skill (equation~\eqref{eq:NLS}) as a function of number of nearest neighbors $n$ retained and the coupling coefficient for geopotential $\alpha_0$, while the other two coefficients are set to one: $\alpha_{1,2} = 1$. 

Other choices for distance function have been explored in the literature, such as Mahalanobis metric~\citep{yates2003technique,stephenson1997correlation}), which is equivalent to performing ZCA ``whitening'', i.e. a procedure that is a rotation of a PCA components plus normalizing the covariance matrix. We do not explicitly compute Mahalanobis distance in this work, although we do actually compute the Euclidean distance {after performing the dimensionality reduction via PCA as will be described below}.
\subsubsection{Dimensionality reduction}\label{sec:DimensionalityReductionMethods}
{When dealing with high dimensional fields, one often encounters the problem of ``curse of dimensionality''. In these cases it is known that Euclidean distance is poorly indicative regarding the similarity of two data points. It could therefore be useful to project the dynamics onto a reduced space. Furthermore, in this way the construction of a catalog is much more efficient from a computational point of view. Here we decided to adopt two different dimensionality reduction techniques, namely a Variational Auto-Encoder (VAE) and Principal Component Analysis (PCA), also known as Empirical Orthogonal Function (EOF) analysis. In a nutshell, PCA is a linear transformation of the original data, where the samples are projected onto the eigenvectors of the empirical correlation matrix~(see  Appendix~\ref{sec:vae_methods}). Usually the projections are made only on the eigenvectors corresponding to the largest eigenvalues, as they are the ones that contain most of the variability of the data. However, this variability may not be entirely relevant to the prediction of heat waves, so this criterion is not necessarily useful. The VAE instead is a non-linear probabilistic projections onto a latent space (usually with Gaussian measure). It consists of probabilistic \emph{decoder} $p_\theta(x|z)$ and probabilistic (stochastic) \emph{encoder} $q_\phi(z|x)$ and the training is performed with the aim of maximize the probability of the data $p(x)$~\citep{doersch2016tutorial} (more details can be found in Appendix~\ref{sec:vae_methods}). Here, we adopt a deep Fully Convolutional Neural Network (FCNN) as an encoder followed by a up-sampling FCNN decoder. Once the data have been projected into the latent space through one of the two methods, the SWG can be built exactly as explained above, simply replacing the fields in high-dimensional space with low-dimensional ones. }


\subsubsection{SWG committor}\label{sec:SWGcommittor}
Due to the necessity of transitioning between Training (TS) and Validation Sets (VS) (see Figure~\ref{fig:analogs}), we require two transition matrices $\mathcal{M}_{\text{tr}}$ and $\mathcal{M}_{\text{va}}$. The former is a straightforward application of what was discussed above, while the latter allows the trajectory starting in the VS to find the appropriate analog in TS. During this procedure (when $\tau=0$) we retain the mean value of temperature over $\tau_c$ days that was already known in the VS (see Section~\ref{sec:coarse}), and re-use it when computing the synthetic label $A(t)$, equation~\eqref{eq:coarsetimeaveraged}. All the remaining entries in $A(t)$ are based on the temperatures corresponding to the subsequent states visited in the training set.  

When $\tau_m=3$ and $T=15$ and $\tau=0$ days, we need to evolve the trajectory only four times, thus each state $s$ can have at most $K ^ 4$ different values of $A$. Therefore, if we denote by $N_a(s)$ the number of trajectories such that $A$ is greater than $a$, the committor function of the state $s$ will be $p(s)=\frac{N_a(s)}{K^4}$. Because we are also interested in larger lead times such as $\tau=15$ days we actually need to start the trajectories earlier, which leads to exponentially growing {computational} trees: with total 9 steps of Markov chain necessary to be applied starting from each day in VS. 
To reduce the computational burden we use Monte Carlo sampling with $10,000$ trajectories at each day from VS rather than exploring the full tree. We do not observe improvements in the skill after refining with more trajectories. 

Due to large number of trajectories we had to parallelise our python code with a numba package. SWG requires computation of a large matrix of nearest neighbors, for which we use kDTree from scikit-learn package and we run it in parallel on 16 CPU cores of Intel Xeon Gold 6142. {Assuming there was no dimensionality reduction performed prior to this operation,} the feature space resulting from $\mathcal{Z}$ field is 1152 dimensional (number of cells) and we run a grid search to find optimal $\alpha_0$ coefficient in~ Eq. \eqref{eq:timeaveraged} for each of the 5 folds. The procedure takes approximately 18 hours when dimensionality reduction (See Appendix~\ref{sec:vae_methods}) is not applied. 

{
Finally, because of coarse-graining (see Section~\ref{sec:coarse}) and the definition of labels (equation~\eqref{areaintegral}) special care has to be made when comparing quality of the prediction of~\eqref{eq:committor} with respect to~\citep{miloshevich22}, where $\mathcal{X}$ was daily. Indeed, the coarse-grained field $\mathcal{X}$ retains information until $\tau_c$ days later due to forward coarse-graining (equation~\eqref{eq:coarse}). Thus for a fair comparison between different coarse-grained committors one must consistently shift the lead time $\tau$ of $\tau_{c_1}-\tau_{c_2}$ to ensure that fields $\mathcal{X}_1$ and $\mathcal{X}_2$ have no information about the future state of the system. In this paper we consider coarse graining time of 3 days. }



\subsubsection{Return Time Plots}\label{sec:returns_methods}
The return time plots are produced using the method described in details in~\citep{Lestang_2018}. Here we will mostly summarize the algorithm. The idea is to estimate the return time of block-maximum (summer) $A(t)$ surpassing a particularly high threshold value $a$. Thus the heatwave definition differs slightly with respect to what is used in the committor definition (Section~\ref{sec:committor}), i.e. we do not restrict $a$ to the 95-th percentile, but instead identify the largest value of $A(t)$ each summer which will correspond to $a$ of that year. The interval (duration of the block) is defined so that heatwave may start in June 1 and end in August 30, so the interval depends on the duration of the event $T$, which for this application may take values other than just 15 days. If $T=30$ days, for instance, the last heatwave may start no later than August 1. Next, the yearly thresholds $a$'s are sorted in the descending order from largest to smallest: $\left\{a_m\right\}_{1 \leqslant m \leqslant M}$, where $M$ is the total number of years. 
With this definition the most extreme return time, is identical to the length of the dataset (in years) under consideration (Table~\ref{tab:datasets}) and the thresholds are ordered $a_1\ge a_2\ge \dots \ge a_M$. The simplest approach is to just to associate the rank of the thresholds $a$ to the inverse return time. 
\begin{equation}\label{eq:block_max}
    r = \frac{M}{\text{rank}(a)},
\end{equation}
 This definition is intuitively reasonable for large $a$ but runs into problems at the other end of the $a$ axis. 

A more precise estimator uses the assumption of Poisson process $P(t) = \lambda \exp\left( -\lambda t \right)$ approximation (which largely affects the small return times), \emph{modified block maximum estimator} in~\citep{Lestang_2018} which reads in our notation as
\begin{equation}\label{eq:return_equation}
    r = - \frac{1}{\log \left(\text{1 - rank}(a)/M \right)},
\end{equation}
where $M$ is the length of the dataset in years and $\mathrm{rank}(a)$ is the order in which the particular year appears in the ordered list of years (by threshold). 

\subsection{Generalized Extreme Value (GEV) fit}\label{sec:gev}

We chose the {\tt{scikit-extremes}} package of ~\citet{skextremes} to perform the Generalized Extreme Value (GEV) fits. The maximal yearly summer $A(t)$ is calculated and then fit using Maximum Likelihood Method estimator (MLE). Because the initial guess for the parameters is important and can have an impact on the shape parameter of the GEV distribution, linear moments of the distribution are used  to obtain those start values. {The GEV function is as follows}
\begin{equation}
    {f(x, c)=\exp \left(-(1-c x)^{1 / c}\right)(1-c x)^{1 / c-1},}
\end{equation}
{where $c$ corresponds to the shape parameter.}
To estimate the confidence intervals the delta method is employed.
Some years in the simulation tend to be cooler and even with negative temperature anomaly $A_{\text{max}}(T)<0$. In fact there is a handful of such $A_{\text{max}}(T)<0$ years in the TS of D100. To better estimate the GEV parameters, we remove such years.


\section{Results}

\subsection{Probabilistic forecasting}\label{sec:forecasting}
The first goal is to compare Stochastic Weather Generator (SWG) and Convolutional Neural Network (CNN) as tools for probabilistic heatwave forcasting. This is done in a framework depicted schematically on Figure~\ref{fig:architecture}.

\subsubsection{Comparisons with CNN}\label{sec:SWGvsCNN}

\begin{figure}
    \centering
    \includegraphics[width=\linewidth]{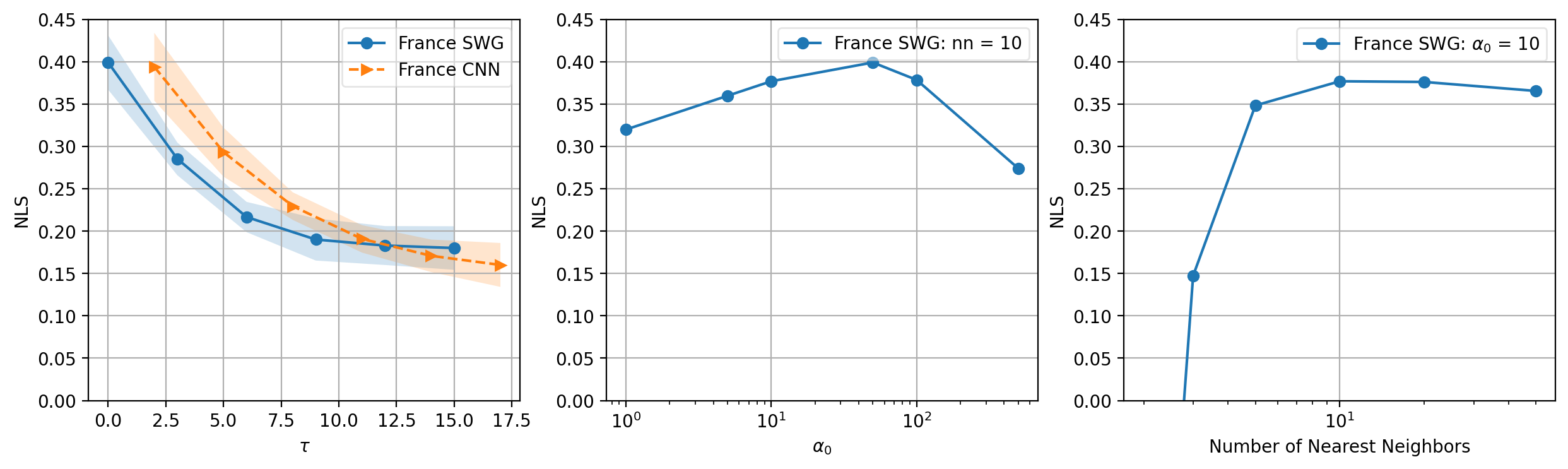}
    \includegraphics[width=\linewidth]{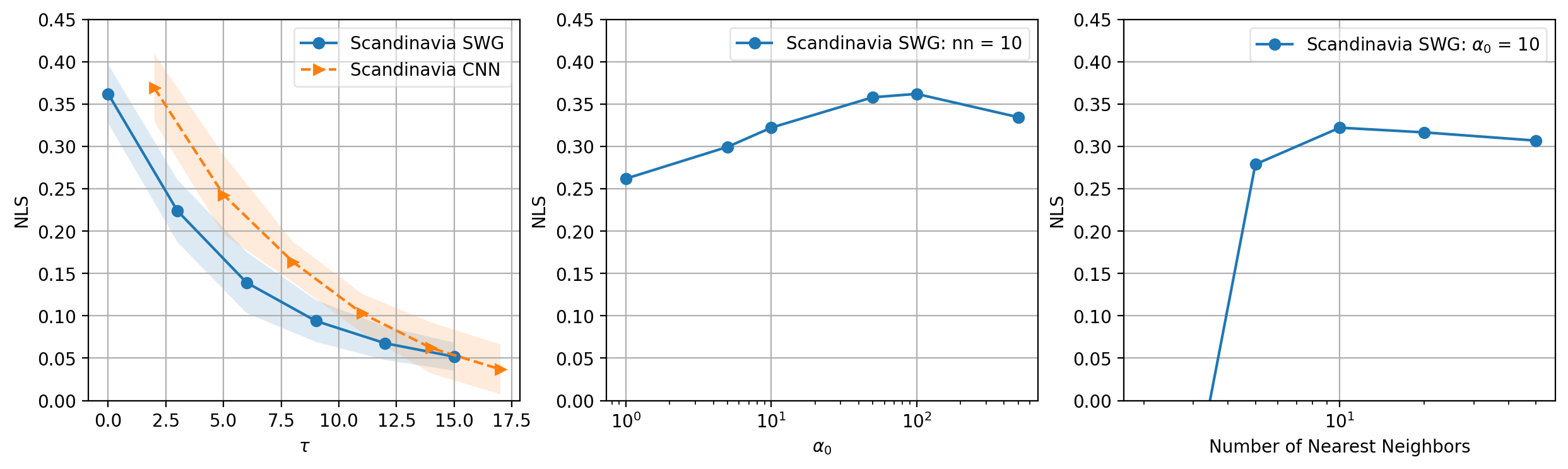}
    \caption{{\bf Basic SWG (blue curve) vs CNN. (orange curve)} All three panels display NLS (equation~\eqref{eq:NLS}) on the $y$ axis. Left panel has $\tau$ on the $x$ axis, central panel has $\alpha_0$ (a hyperparameter of SWG, see equation~\eqref{eq:metric}) on the $x$ axis and right panel has $n$-nearest neighbors (also hyperparameter of SWG) on the $x$ axis. {On the central and the right panels the choice for $\tau = 0$ was made}. The dots show data points corresponding to the mean of the cross-validation, whereas the thickness of the shaded area represents two standard deviations. These conventions will be re-used in the subsequent figures. 
    }
    \label{fig:tau_simple_analog}
\end{figure}

Here we plot the NLS (equation~\eqref{eq:NLS}) comparing the predictions of CNN (orange curve) vs SWG (blue curve) described in Sections~\ref{sec:CNN} and~\ref{sec:analog_method} for France (top panels) and Scandinavia (bottom panels) respectively on Figure~\ref{fig:tau_simple_analog}. 
{The comparison is made on D500 (see Table~\ref{tab:datasets}) but similar results are obtained considering shorter dataset (see Appendix~\ref{sec:committorD100}).} The left panels are plotted as a function of lead time $\tau$ and just like in~\citep{miloshevich22} we see a downward trend in NLS due to chaotic nature of the atmosphere (the predictability diminishes with time).  As was described in Section~\ref{sec:state_space_metric} parameters $\alpha_0$ and the number of nearest neighbors of SWG were tuned based on cross-validation which allows us to provide error bars (shading around the curves). We find that using 10 nearest neighbors is almost always optimal (right panels), whereas the optimal values of $\alpha_0$ (middle panels) tend to be of the order $50-100$. Later we shall see that $\alpha_0$ also may depend on the preprocessing the data (such as dimensionality reduction).

We are following identical procedures of data processing for both SWG and CNN
(Section~\ref{sec:training_validation}). However, because of the coarse-graining in SWG the comparisons are only appropriate, if we shift orange curve by $\tau_c-1=2$ days ahead (increasing $\tau$).  {The latter adjustment is explained in Section~\ref{sec:SWGcommittor}}. The Figure~\ref{fig:tau_simple_analog} shows that in France at $\tau = 2$ days NLS reaches approximately $0.40\pm 0.04$ for CNN and an estimation of $0.30 \pm 0.02$ for SWG. In Scandinavia (bottom panel) we have similar results: approximately $0.37\pm 
 0.03$ for CNN vs only $0.25 \pm 0.02$ for SWG. Large lead times $\tau$ result in lower predictive skill due to chaotic nature of the atmosphere. The behavior of the curve leads to the conclusion that CNN predicts heatwaves better up to 10 days before the event at which point the CNN and SWG skills are not statistically distinguishable. In case of France the regime of large $\tau$ is mostly dominated by predictability due to soil moisture~\citep{miloshevich22}, whereas in Scandinavia this long-term information is absent thus NLS converges to near-zero values at large $\tau$. It is generally known~\citep{felsche23}, that it is heatwaves in Southern Europe that are preceded by the negative anomaly of soil moisture, rather than Northern Europe, which is consistent with what we see here. The reasoning we provide is that in France the soil may actually be in the regime of strong lack of humidity in summer which significantly amplifies the soil-atmosphere feedbacks. 

Looking at the optimality of the parameter $\alpha_0$, which controls the importance of geopotential in the metric (equation~\eqref{eq:metric}), we see that  for France it is of order $\alpha_0 \sim 50$, an order of magnitude higher than what we expected $\alpha_0 = 1$ based on weighting by the number of relevant grid points. In Scandinavia the optimal value is not too different, of order $\alpha_0 \sim 100$, but the dependence is weaker. The reason for this could also be the fact that soil moisture does not contribute much to heatwave predictability in that region, thus limiting its usefulness in the metric (equation~\eqref{eq:metric}). 

The benchmarks indicate that CNN outperforms SWG independent of the area of interest (France or Scandinavia) or even dataset length (Figure~\eqref{fig:tau_simple_analog_D100}). 


\subsubsection{Dimensionality reduction}

\begin{figure}
       \includegraphics[width=\linewidth]{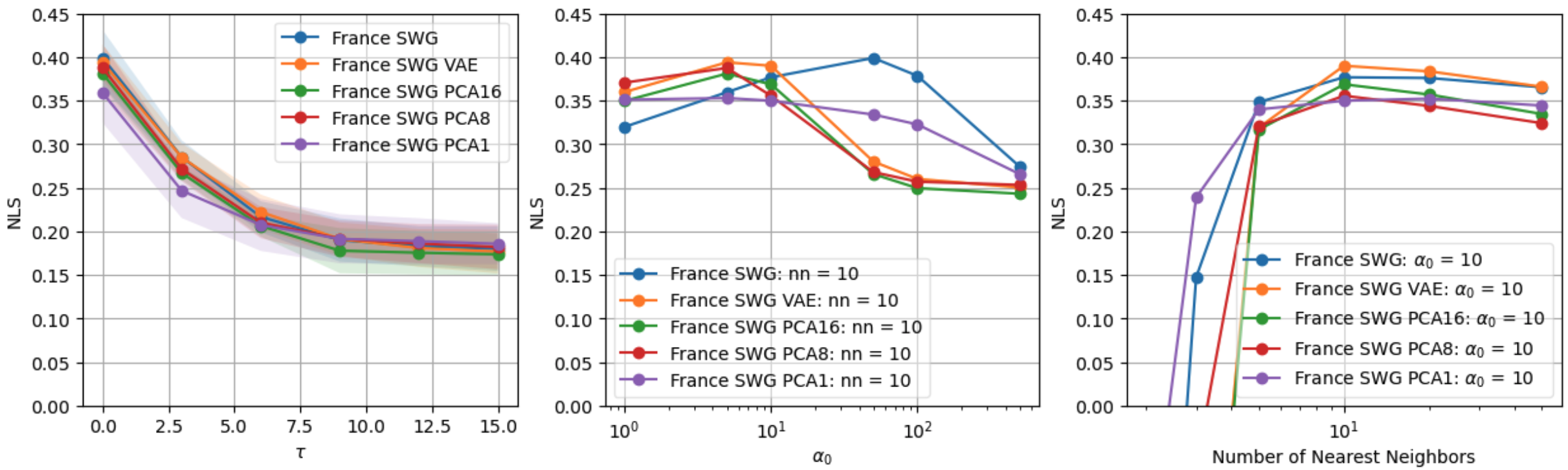}
  
      \caption{{\bf Basic SWG vs VAESWG:}  On the $y$ axis we have NLS (equation~\eqref{eq:NLS}) as a function of lead time $\tau$ and hyper-parameters of SWG (see caption of Figure~\ref{fig:tau_simple_analog}). SWG is indicated by the same (identical) blue curve as in Figure~\ref{fig:tau_simple_analog} while orange and green curves correspond to VAESWG where geopotential was passed through two different autoencoders (equation~\eqref{eq:vae_ZG}) (orange and green curves).  }
    \label{fig:tau_analog_vae_f}
\end{figure}

\begin{figure}
       \includegraphics[width=\linewidth]{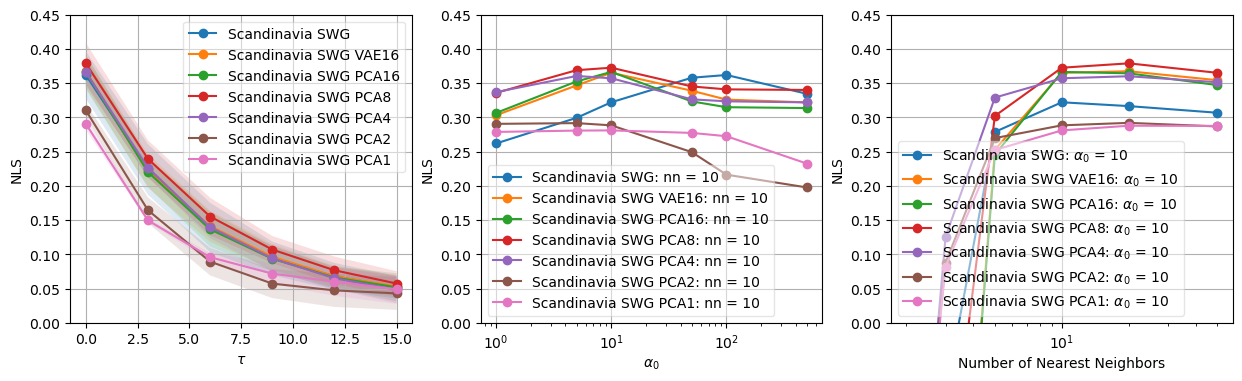}
  
      \caption{{\bf Basic SWG vs VAESWG:}  On the $y$ axis we have NLS (equation~\eqref{eq:NLS}) as a function of lead time $\tau$ and hyper-parameters of SWG (see caption of Figure~\ref{fig:tau_simple_analog}). SWG is indicated by the same (identical) blue curve as in Figure~\ref{fig:tau_simple_analog} while orange and green curves correspond to VAESWG where geopotential was passed through two different autoencoders (equation~\eqref{eq:vae_ZG}) (orange and green curves).  }
    \label{fig:tau_analog_vae_s}
\end{figure}


{We now move on to the question of projecting the underlying space to latent dimension as described in Sec.~\ref{sec:DimensionalityReductionMethods}.} Three approaches are considered, one which does not involve any dimensionality reduction, one where a deep Fully Convolutional Neural Network (FCNN) is used as an encoder followed by a up-sampling FCNN decoder and one, where instead of autoencoder we use scipy package PCA projections\footnote{Principal Component Analysis (PCA) is known as Empirical Orthogonal Function (EOF) analysis in the field of geosciences.}. 
{We plot the results on Figures~(\ref{fig:tau_analog_vae_f},\ref{fig:tau_analog_vae_s}) for France and Scandinavia, respectively. There, simple SWG trained without dimensionality reduction is represented via a blue curve, autoencoder (latent dimension $p=16$) via orange and the other curves represent PCA for different latent dimensions $p$. For France we see no statistically significant differences even in the drastic case where the latent dimension consists of the two largest principal components only\footnote{{Note that we speak of two largest principal components of the geopotential, but SWG also accepts local temperature and soil moisture as input predictor variables.}}. This is probably due to the strong predictability power of soil moisture field (as clearly shown in~\citet{miloshevich22}). Indeed, the results for Scandinavia (which is less affected by soil moisture) shown in Fig~(\ref{fig:tau_analog_vae_s}) display a degradation when $p<4$ while for $p\ge4$ all methods provide essentially the same score. Interestingly, the highest score (although within the error bars) is obtained for $p=8$, suggesting that an optimal dimension may exist.   
}

The computation of analogs (equation~\eqref{eq:metric}) is rather difficult in high dimensions and with datasets larger than $1000$ years becomes unfeasible, because we also have to perform tuning of $\alpha_0$ and the number of nearest neighbors. This is much more efficient if the dimension of the space has been successfully reduced, with a method such as autoencoder. However as can be seen from  Figures~(\ref{fig:tau_analog_vae_f},\ref{fig:tau_analog_vae_s}) the benefits from using complicated multi-layer architecture are limited in this case, given the fact that the green curve, corresponding to PCA is almost the same (within the error bars) as the one obtained with the autoencoder. One notable difference between the cases with dimensional reduction and without is that the value for $\alpha_0$ that is optimal $\alpha_0 = 5$ is an order of magnitude smaller than the one we obtain on Figure~\ref{fig:tau_simple_analog}, corresponding to SWG learned in real space. 

These conclusions should be taken in the context of the predictability of temperature using geopotential, which have rather linear Gaussian statistics. It could be that autoencoders are more useful for more complex fields such as precipitation. 


\subsection{Sampling extremes}\label{sec:sampling}
From now on our goal will be to use Stochastic Weather Generator (SWG) and evaluate the quality of the statistics for the extremes it can produce. In contrast to the Section~\ref{sec:forecasting} we will consider heatwave extremes as large as one in 7000 years events.

\subsubsection{Computing return times}\label{sec:return_results}

\begin{figure}[ht]
    \centering
    \begin{subfigure}[b]{.48\textwidth}
        \centering
        \includegraphics[width=\textwidth]{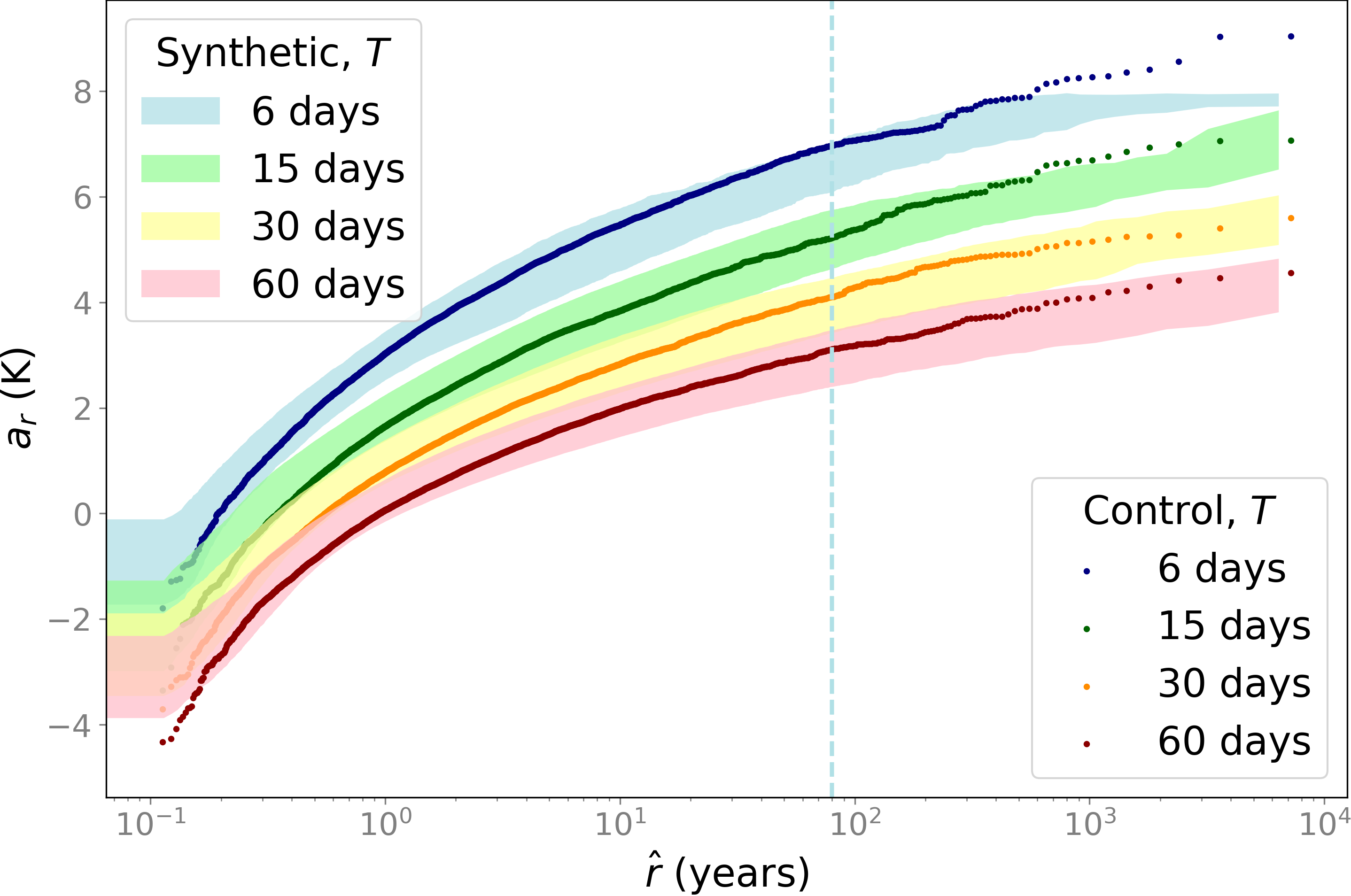}
        \caption{France return times}
         \label{fig:FranceReturns}
    \end{subfigure}
    \begin{subfigure}[b]{.48\textwidth}
        \centering
        \includegraphics[width=\textwidth]{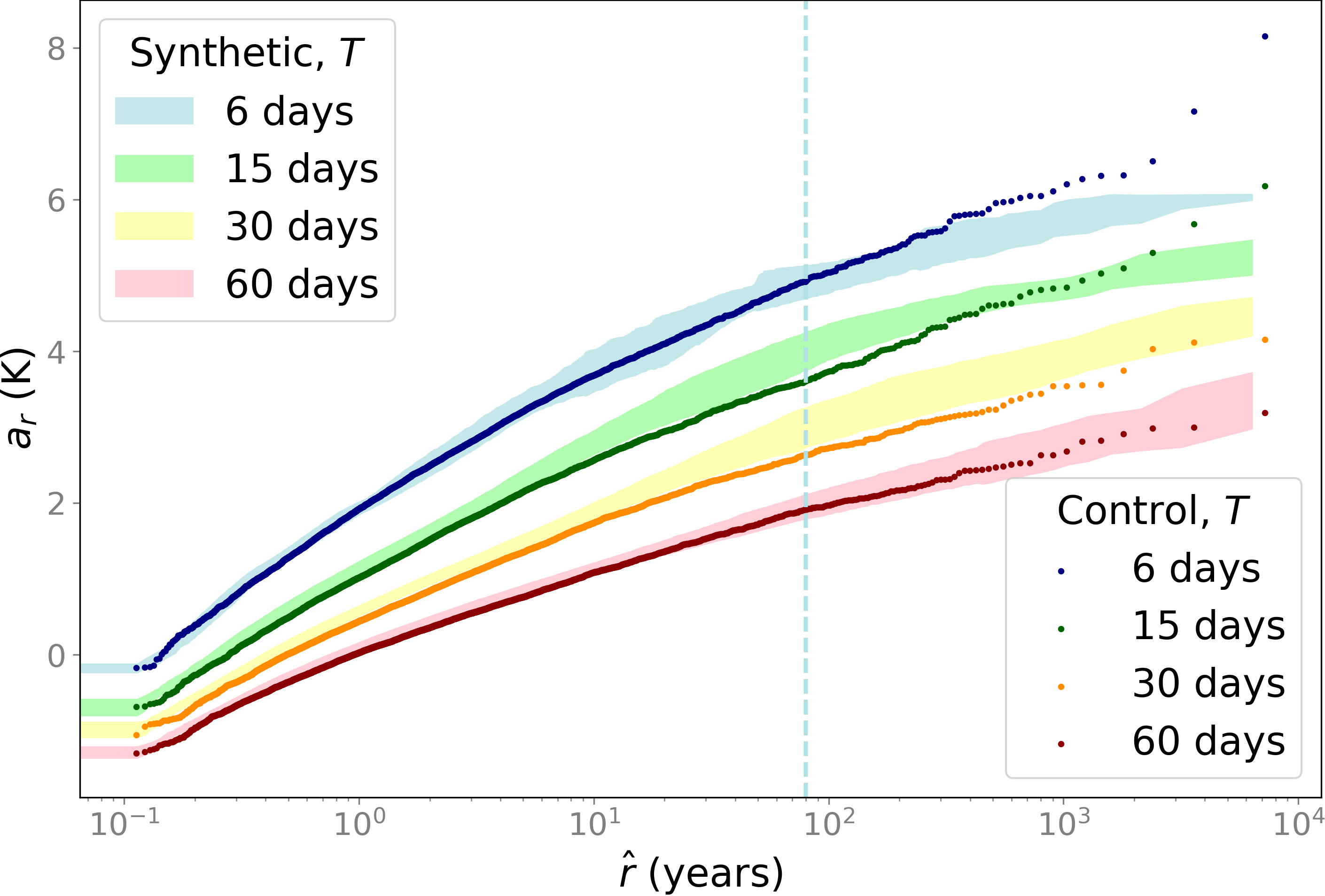}
    \caption{Scandinavia return times}
    \label{fig:ScandinaviaReturns}
     \end{subfigure}
    \caption{Return time plot for (a) France and (b) Scandinavia  heatwaves using analogues of North Atlantic and Europe and the method based on equation~\eqref{eq:return_equation}. Here we use parameters $\alpha=1$ (default), Number of nearest neighbors $n=10$, the analogs are initialized on June 1 of each year (using the simulation data) and then advanced according to the Algorithm~\ref{alg:SWG}. The trajectory ends at the last day of summer. Each trajectory is sampled 800 times providing much longer synthetic series and thus estimating longer return times. Return times are computed for $T=\{6,15,30,60\}$ day heatwaves (indicated on the inset legend), with dots corresponding to the statistics from the control run (D8000, see Table~\ref{tab:datasets}), while shaded areas correspond to the bootstrapped synthetic trajectory: the whole sequence is split into 10 portions which allows estimating the mean and variance. The shading corresponds to mean plus or minus one standard deviation.  }
    \label{fig:Returns}
\end{figure}

We aim to estimate extreme return times from shorter sequences and compare them to the control run D8000 (Table~\ref{tab:datasets}). To this end we use SWG trained on D100 (Table~\ref{tab:datasets}) and initialized with $\alpha_0=1$ and 10 nearest neighbors, parameters that we do not optimize in contrast to Section~\ref{sec:forecasting}. The generated synthetic sequences are plotted for France on Figure~\ref{fig:FranceReturns} and for Scandinavia on Figure~\ref{fig:ScandinaviaReturns}. The curves were obtained using the method described in Section~\ref{sec:returns_methods}, so that the analogs are initialized in June 1 of each year (using the simulation data) and then advanced according to the Algorithm~\ref{alg:SWG}. The trajectory lasts up to synthetic August 30. 

The main conclusion from Figure~\ref{fig:FranceReturns} is that the analog method is rather well-suited for this task: not only do synthetic trajectories match return times of the reference real trajectory (80 years of analogs) but they also provide correct estimates of the returns of a much longer trajectory (8000 years) at a fairly low computational footprint, compared to running such a trajectory. For instance, for $T=15$ day heatwaves the most extreme event with return time of 7200 years has anomaly $7.07 K$ which is well within the error bars of SWG estimate $7.02 \pm 0.46 K$. We can see that the generalization happens consistently, except for the extremes of 6 day heatwaves, where we have sampling issue since $\tau_m = 3$ days. The procedure is repeated for Scandinavia on Figure~\ref{fig:ScandinaviaReturns}. The results are {generally consistent}, although we observe more deviations from the prediction of SWG.  For example, heatwaves of length $T=15$ days and return times of order 2-3 thousand years as well as the longest return times for $T=30$ days tend to be systematically underestimated.
{Note, however, that these events are rare even in the control run and therefore their thresholds have large uncertainties. Excluding these extreme events, the predictions are almost always in agreement with the return times calculated on the long run and when they are not, the relative error is smaller than $15\%$.}

\begin{figure}[ht]
    \centering
    \begin{subfigure}[b]{.45\textwidth}
        \centering
        \includegraphics[width=\textwidth]{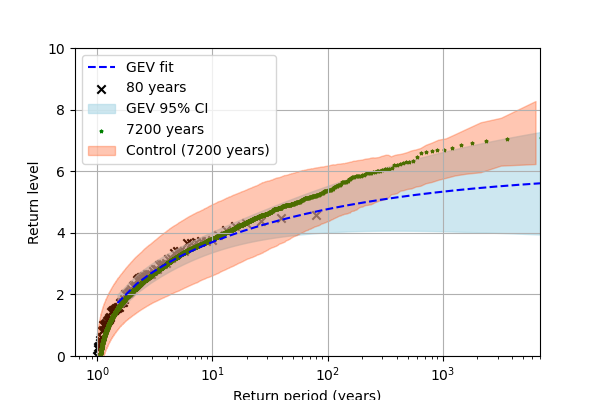}
        \caption{France return times}
         \label{fig:FranceReturnsGEV}
    \end{subfigure}
    \begin{subfigure}[b]{.45\textwidth}
        \centering
        \includegraphics[width=\textwidth]{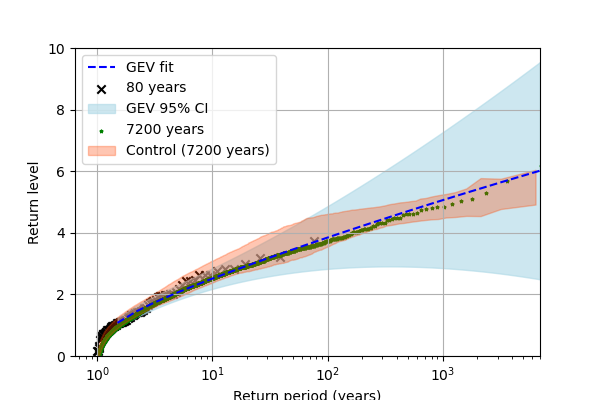}
    \caption{Scandinavia return times}
    \label{fig:ScandinaviaReturnsGEV}
     \end{subfigure}
    \caption{Return time plot for (a) France and (b) Scandinavia $T=15$ day heatwaves using the method based on equation~\eqref{eq:block_max}. On the $y$ axis $a$ anomalies are plotted in $K$ and on the $x$ axis years. The black dots correspond to the most extreme heatwaves in 80 years of D100. The blue dashed line corresponds to the GEV fit performed on 80 years of D100 (minus the ones which have negative $A(t)$ maxima). The 95 percent confidence intervals are indicated via blue shade.  The synthetic  time series generated via SWG and identical to the green shade on Figure~\ref{fig:Returns} are plotted using orange shade, except that we chose to shade two standard deviations, rather then one for consistency. The 7200 years control run (identical to green dots on Figure~\ref{fig:Returns}) are plotted using green dots.  }
    \label{fig:ReturnsGEV}
\end{figure}

Since for this type of risk assessments Extreme Value Theory (EVT) is often employed, we compare our approach to the Generalized Extreme Value (GEV) function fits obtained using package developed by~\citep{skextremes} from the 79 extreme $T=14$ day heatwaves in the 80 years of D100 (our training data) on Figure~\ref{fig:ReturnsGEV}. For details on why only 79 heatwaves are chosen, see Section~\ref{sec:gev}. 

Generally speaking, GEV fit performs worse than SWG; in France it underestimates the severity of extremes, while in Scandinavia it produces confidence intervals that are very large. {While the GEV fit is also consistent with the extremes we observe in the control run it provides much looser confidence intervals. On the other hand, SWG tends to shadow more closely the extremes of the long control simulation.}
This could be ascribed to the hypothesis that there is an upper bound for temperature extremes (e.g.~\citet{zhang23}). Parameters of the GEV fit yield uncertainties (especially the shape parameter) which can be very large when the available data is limited (here: 80 years). This leads to wide uncertainties for return periods that are larger than the training length, although in case of Scandinavia the return levels of the long control simulations do fall within the confidence intervals of the GEV fits, albeit not \emph{on} the GEV fit. On the other hand, the SWG simulations are close to intrinsic properties of temperature, driven by the predictors we use. This explains why the SWG simulations follow the long control run, with a relatively narrow confidence interval, which does not increase with return periods.


We do not control for $\alpha_1$ and $\alpha_2$ to give us insight into which variable matters more in this aspect (i.e. temperature or soil moisture), however since soil moisture does not have impact on heatwaves on Scandinavia, one could reason that it is temperature in that case. In other words, to produce reliable return time plots the Euclidean metric should not only take circulation patterns into account but also the temperature and soil moisture in some cases.
A side criticism of our approach is that we have chosen value $\alpha_0 = 1$, which seems arbitrary in view of the discussion in Section~\ref{sec:SWGvsCNN} on the optimality of $\alpha_0 \sim 50$ for performing the probabilistic forecast. {However, it should be noted that having changed the task to be performed, there is no longer a guarantee that $\alpha_0\sim 50$ is still an optimal value (and it is not, as shown in Appendix~\ref{sec:returnsalpha50}).}

\subsubsection{Teleconnection patterns for extremes}

\begin{figure}[ht]
    \centering
    \begin{subfigure}[b]{.4\textwidth}
        \centering
        \includegraphics[width=\textwidth]{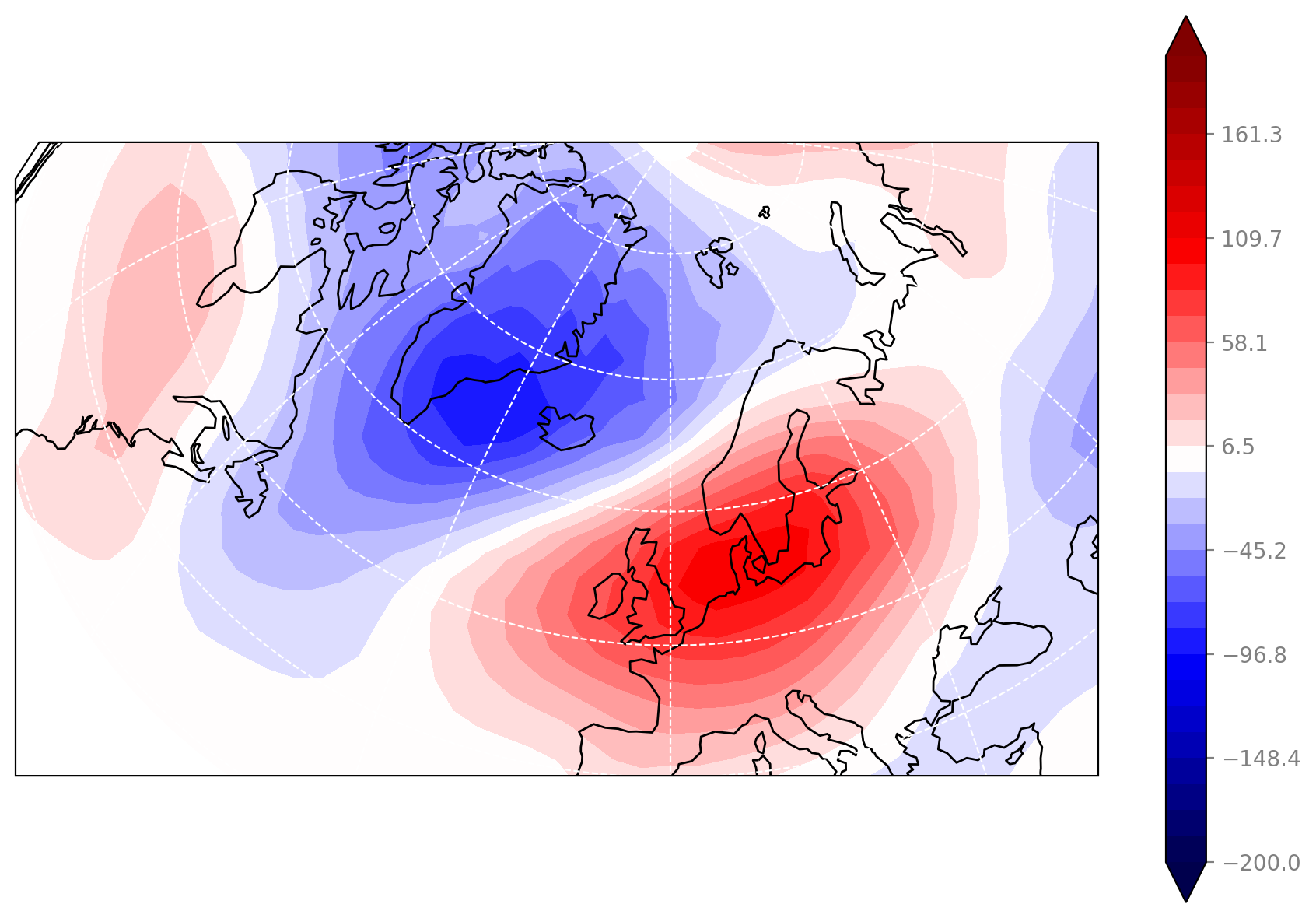}
        \caption{France 80 years}
         \label{fig:France100yrs}
     \end{subfigure}
     \hfill
    \begin{subfigure}[b]{.4\textwidth}
        \centering
        \includegraphics[width=\textwidth]{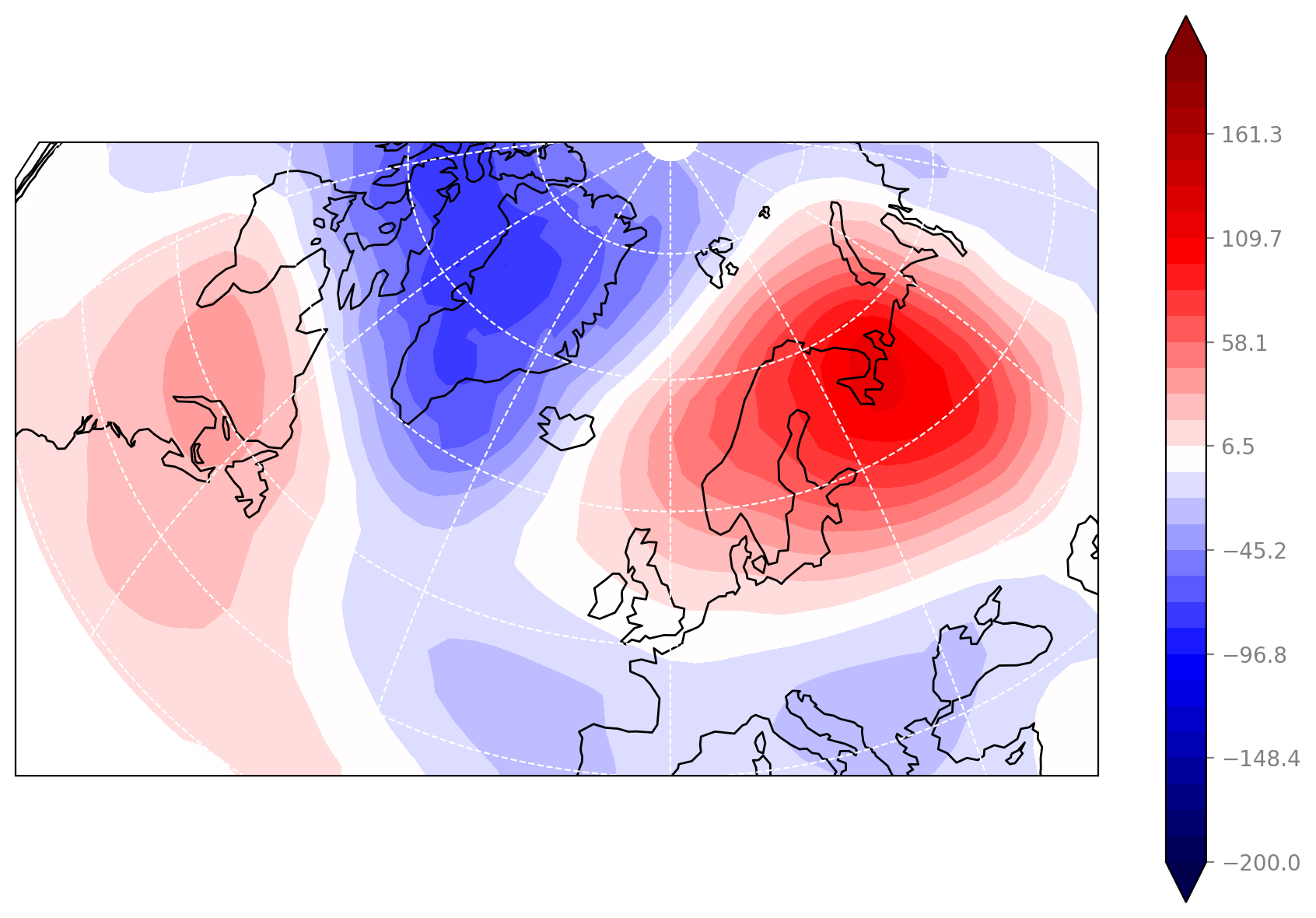}
        \caption{Scandinavia 80 years}
        \label{fig:Scandinavia100yrs}
    \end{subfigure}
     \hfill
     \hfill
     \begin{subfigure}[b]{.4\textwidth}
        \centering
        \includegraphics[width=\textwidth]{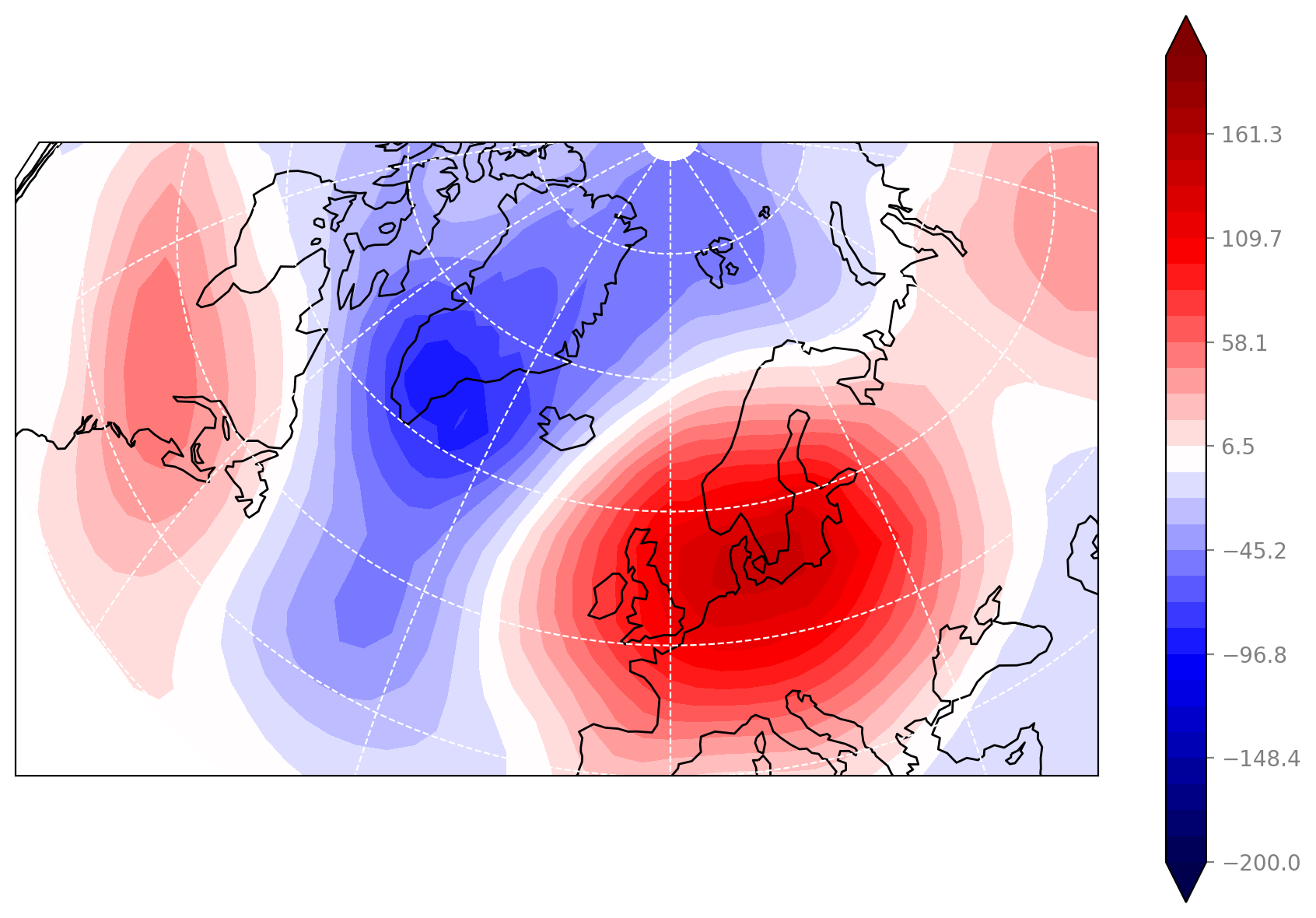}
        \caption{France control run}
        \label{fig:France7200yrs}
    \end{subfigure}
    \hfill
    \begin{subfigure}[b]{.4\textwidth}
        \centering
        \includegraphics[width=\textwidth]{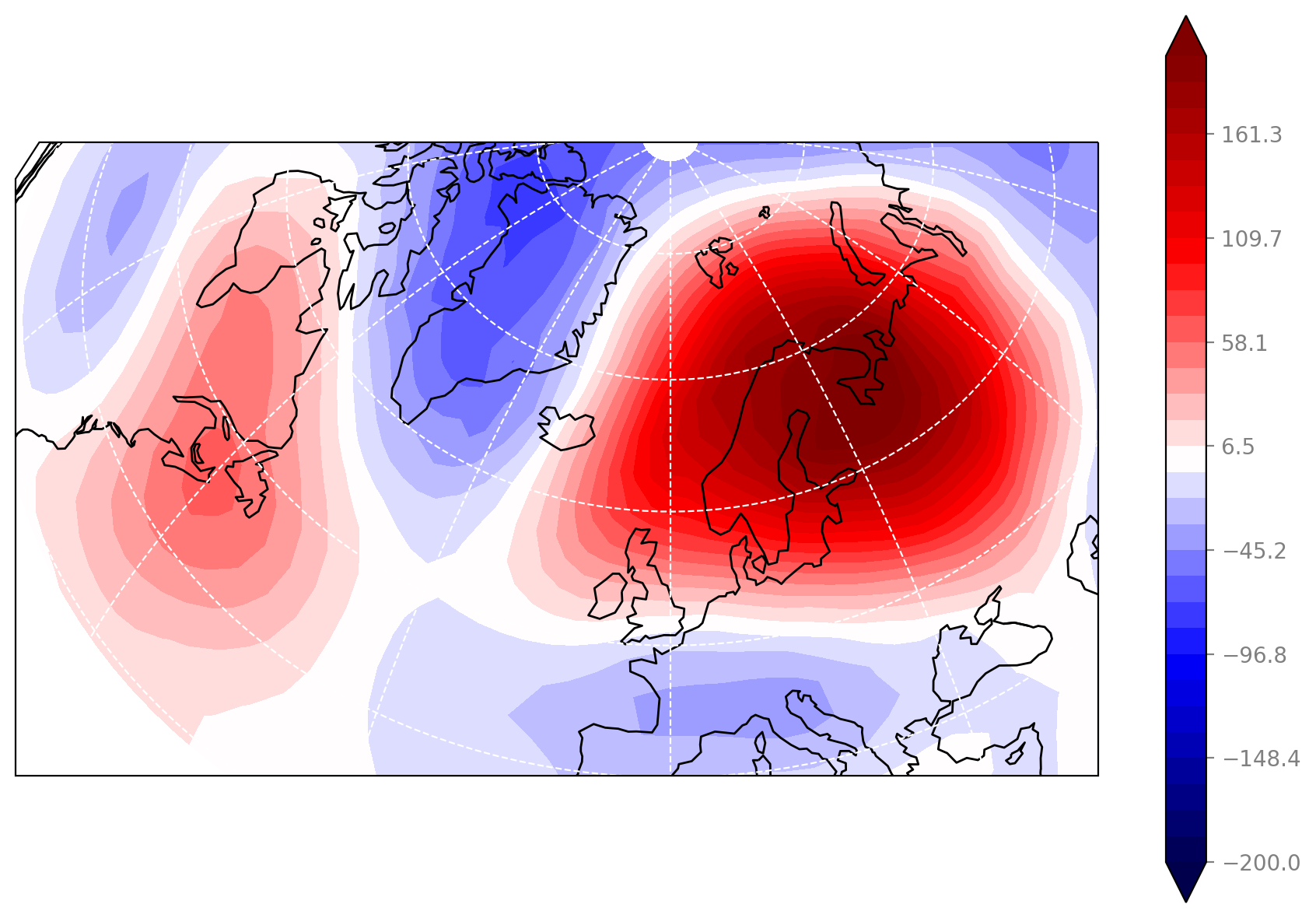}
        \caption{Scandinavia control run}
        \label{fig:Scandinavia7200yrs}
    \end{subfigure}
    \hfill
    \begin{subfigure}[b]{.4\textwidth}
        \centering
        \includegraphics[width=\textwidth]{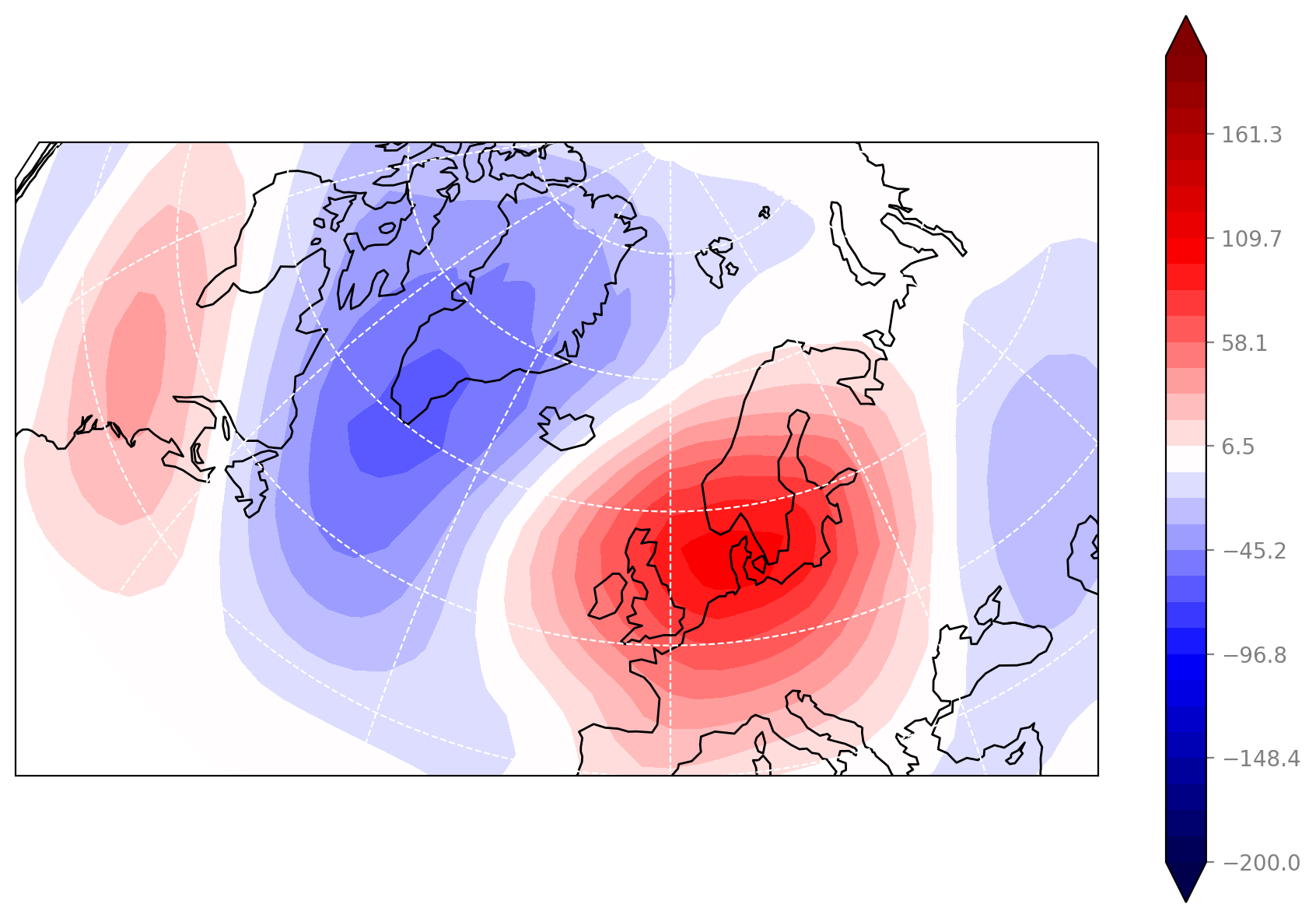}
        \caption{France synthetic SWG}
        \label{fig:FranceSWG}
    \end{subfigure}
    \hfill
    \begin{subfigure}[b]{.45\textwidth}
        \centering
        \includegraphics[width=\textwidth]{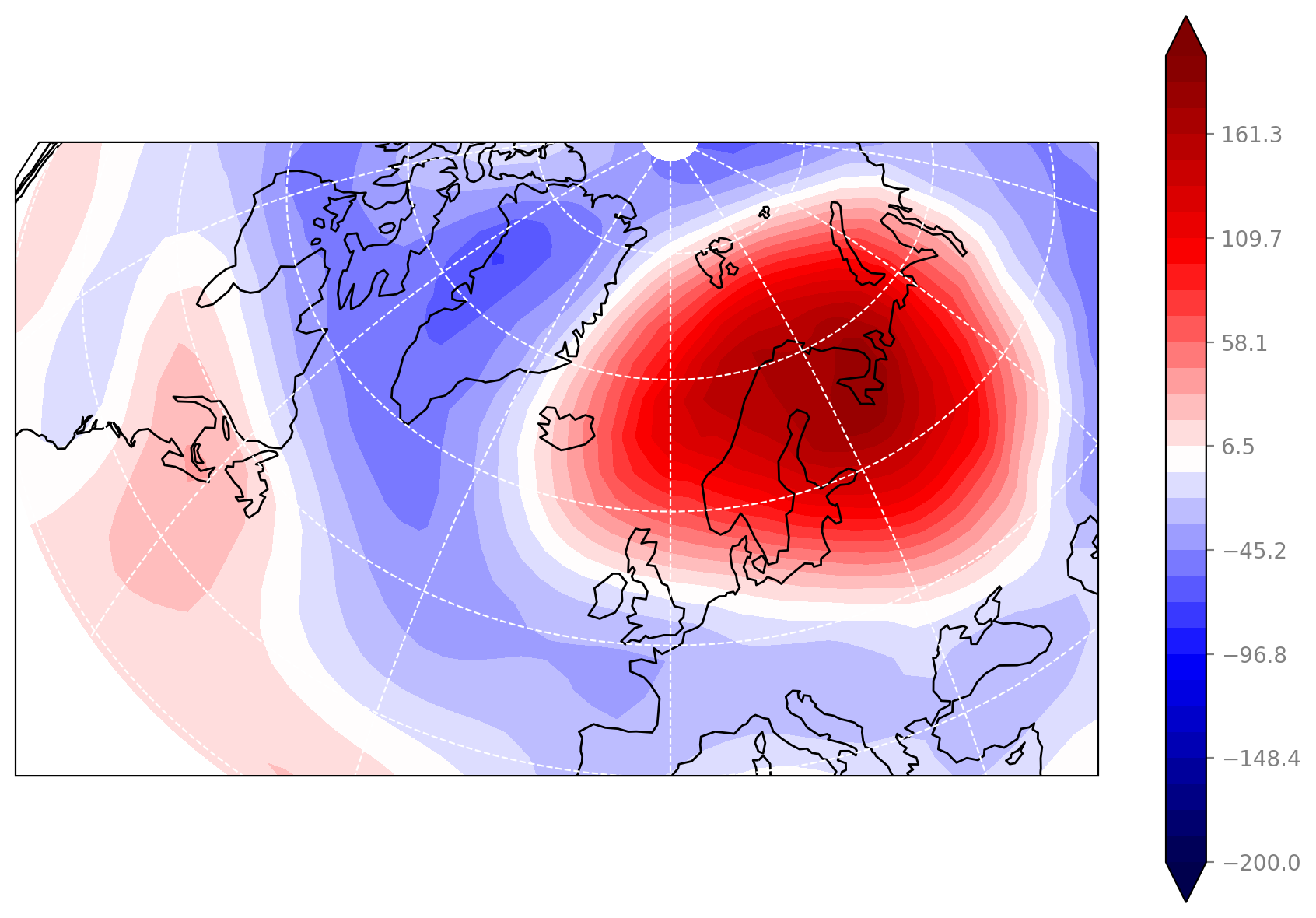}
        \caption{Scandinavia synthetic SWG}
        \label{fig:ScandinaviaSWG}
    \end{subfigure}
    \caption{Composite maps of geopotential height (meters) {anomalies at 500 hPa for heatwave in France and Scandinavia ($T=15$ days). (a) forward 3 day running mean at $\tau = 0$, i.e. the heatwave onset, {a composite of the 10} most extreme France heatwaves in 80 year long dataset with the threshold $\approx 4$ K. {(b) Same as (a) but for Scandinavia heatwaves.} (c) Composite for the control run with a collection of France heatwaves above the threshold $4$ K. (d) Same as (c) but for Scandianvia heatwave (e) Composite for the synthetic run performed by SWG above the same threshold for France heatwaves. (f) same as (d) but for Scandinavia heatwave.} }
    \label{fig:teleconnections}
\end{figure}

We use synthetic trajectories generated by SWG trained on only 80 years (from D100, see Table~\ref{tab:datasets}) to estimate extreme teleconnection patterns that were never observed in that run. Note that we do not use autoencoder to generate new circulation patterns. Instead SWG generates new sequences from the already existing ones which results in new values of $A(t)$ (equation~\eqref{eq:timeaveraged}) since it is a running mean over temperature series. This is precisely the recipe we have followed in Section~\ref{sec:return_results}. To validate the teleconnection patterns thus obtained, we compare them to the control run (D8000). The results are plotted in Figure~\ref{fig:teleconnections}. 

First, we identify the {10 most extreme heatwaves} in D100 and plot the composite on their first three days ($\tau = \{0,1,2\}$ days) since the onset on Figure~\ref{fig:France100yrs} \footnote{this is inline with $\tau_m = 3$ day coarse-graining approach we have taken for SWG inputs as explained in Section~\ref{sec:coarse}}. The threshold for this event happens to be approximately $4 K$. The pattern consists of anticyclonic (in red) and cyclonic (blue) anomalies (the word ``anomalies'' will be suppressed in what follows for brevity). 

Next, we compare this lack of data regime {(essentially 10 events)} to the composite map that can be computed using a 7200 years long control run on Figure~\ref{fig:France7200yrs} using the same $4 K$ threshold and plotting the first three days ($\tau = \{0, 1, 2\}$ days) of the composites. In contrast to the previous situation, this leaves us with many more events satisfying the constraint.  The picture changes {somewhat}, most notably, {the cyclone and anticyclone} become more pronounced, while the other patterns {maintain relationship consistent with a Rossby wave pattern}. This is in line with the understanding of teleconnection patterns expected in European heatwaves.  This pattern is also consistent with wave-number 3 teleconnection for heatwaves in France obtained in~\citet{Ragone21} and ~\citet{miloshevich2023robust} computed for 1000 year long sequence. 

Finally, we compare both figures with a very long synthetic SWG run on Figure~\ref{fig:FranceSWG}. The resulting teleconnection pattern {is again similar to the one on} ~\ref{fig:France7200yrs}. {Overall, the advantage offered by SWG in this case is rather small and mostly has to do with the shape of the anticyclone over Northern Europe which in the short D100 composites appears more narrow}. 

We repeat the exact same steps but for the heatwaves in Scandinavia, starting from the {10} most extreme events in D100 on Figure~\ref{fig:Scandinavia100yrs} to the composites resulting from synthetic SWG run on Figure~\ref{fig:ScandinaviaSWG}. 
{In this case SWG performs better in that it is able to capture the magnitude of the anticyclone over Scandinavia more accurately (compared to Figure~\ref{fig:Scandinavia7200yrs}) than D100 composite, although the other features are hard to distinguish.}

\section{Discussion}
We have systematically compared performance of Stochastic Weather Generator (SWG) and Convolutional Neural Network (CNN)~\citep{miloshevich22} tasked to predict the occurrence of prolonged heatwaves over two European areas using simulation data from intermediate complexity climate model Plasim. We have also studied the ability to sample extreme return times and composite maps via the method of SWG. 

In addition to the study of heatwaves in France considered in~\citep{miloshevich22} we also apply our methodology to heatwaves in Scandinavia. This area has different hydrology which impacts long-term predictability. In order to achieve better predictive ability with SWG we took the version developed earlier~\citep{yiou2014anawege}, which previously used only analogs of circulation, predictors we refer to as \emph{global}, and upgraded it  with additional predictors such as temperature and soil moisture integrated over the area of interest (where heatwaves occur), which we refer to as \emph{local}. This was done because of the ample evidence of importance of slow drivers such as soil moisture~\citep{zeppetello2022physics}. Consequently, we had to use appropriate weights in the definition of Euclidean distance to control the importance of global vs local predictors. The benchmarks of CNN vs SWG were obtained based on Normalized Logarithmic Score (NLS) which is particularly well suited for studying rare events~\citep{benedetti10}.  The conclusion is that CNN outperforms SWG in probabilistic forecasting, although this is more evident for particularly large (hundreds of years long) training sets, which would not be available in the observational record. 
This is in line with the statement of~\citep{miloshevich22} on the convergence of such methods in deep learning that requires massive datasets. These conclusions could be interesting for extreme event prediction but are also relevant for future developments of geneological rare event algorithms that aim to resample heatwaves based on the forecasts provided by either CNN or SWG (in the approach similar to~\citep{lucente2022coupling}). 

We have studied the performance of SWG in relation to estimation of large return times and extreme teleconnections. This is achieved by generating synthetic trajectories of extensive length based on the initial training data of a short 80 year Plasim run. We have shown that out-of-the-box (without hyperparameter optimization) implementation of SWG, which takes the three relevant fields as input for the Euclidean metric, was capable of reproducing the return times of the much longer control run (7200 years) for a range of heatwave durations: 15 day, 30 day and 90 day ones. These SWG calculations shadow more closely the control run (with smaller variance) than the Extreme Value Theory (EVT) estimates which are typically used to address similar questions. However, the version of SWG whose weights have been optimized for a different task of conditional (intermediate range to subseasonal) probabilistic prediction tends to underestimate the return times slightly. Nevertheless, its synthetic teleconnection patterns have qualitatively accurate features. 
This is an important result in view of works such as~\citep{yiou:hal-03921111} that rely on SWG to estimate the risks of the extremes. 

In the hopes of improving the performance of SWG we have considered two dimensionality reduction techniques to project the geopotential to the smaller latent space. These techniques involve traditional (Empirical Orthogonal Functions) EOFs and Variational Autoencoders (VAEs). We found that while improving the efficiency at which the analogs are computed, the dimensionality reductions did not modify the predictive skill, which leads us to suggest EOF as a superior dimensionality technique for this task. However, it is possible that VAE could still be useful for other types of extremes, such as precipitation, where applications of traditional analog method is more controversial. We leave such studies for the future.  {When looking for optimal dimensionality of the latent space we found that few 500 hPa geopotential EOF components coupled with local temperature and soil moisture are sufficient to obtain optimal forecasting skill for SWG, which, however, as stated earlier, falls short of CNN prediction which takes the full fields as in input.} 

Another possibility of extending this work would be comparing SWG to other modern tools for learning propagators, for instance, simple U-Net architectures or a corresponding Bayesian Neural Network (BNN) \citep{thuerey20,thuerey2021pbdl}, which provides uncertainty of the prediction thus a kind of propagator. Many other types of architectures are possible, however, given overall simplicity of SWG when training climate model emulators with small datasets SWG should be treated as an indispensable baseline method to justify the use of complex architectures. Fortunately, we have supported this paper with all the necessary code and documentation for straightforward implementation of SWG that can be applied to other projects. 

As stated above, another interesting potential application for SWGs and other statistical methods such as deep learning is rare event algorithms~\citep{Lucente2022committor}. 
Rare event algorithms have been used in past to sample heatwaves~\citep{Ragone18,Ragone21} and could potentially allow to sample other extreme events~\citep{webber19} with expensive high fidelity models at a cheaper cost. Data driven approaches could be used to improve such rare event simulations. This is because knowing the probability of an event gives a prior information to the rare event algorithm~\citep{Chraibi21}. A recent work~\citep{jacques2022data} compares SWG and CNN in computing importance functions for two simple Atlantic meridional overturning circulation models. We believe that providing similar benchmarks is important, thus in this manuscript we concentrate on the comparisons between deep learning and analog forecasting.


Finally, the results here were obtained based on dataset generated by intermediate complexity climate model, PlaSim with the ocean that was driven using stationary climatology and with relatively coarse resolution.  As other modes of long-term predictability come from the ocean the next study should investigate the same questions based on higher fidelity model with better parametrisations. In particular, the question of transfer learning should be addressed, i.e. how to properly generalize out-of-sample by pretraing on a simpler models and fine tuning on more complex datasets, such as CESM or reanalysis.

\section*{Data availability statement}
{Data for this study is available from \href{https://zenodo.org/records/10102506}{https://zenodo.org/records/10102506}.}

\section*{Code availability statement}

The coding resources for this work, such as the python and jupyter notebook files, are available on a GitHub page \href{https://github.com/georgemilosh/Climate-Learning}{https://github.com/georgemilosh/Climate-Learning} and is part of a larger project at LSCE/IPSL ENS de Lyon with multiple collaborators working on rare event algorithms.

\section*{Funding statement}

This work was supported by the ANR grant SAMPRACE, project ANR-20-CE01-0008-01 and the European Union’s Horizon 2020 research and innovation programme under grant agreement No. 101003469 (XAIDA). This work has received funding through the ACADEMICS grant of the IDEXLYON, project of the Universit\'e de Lyon, PIA operated by ANR-16-IDEX-0005.

\section*{Acknowledgement}

We acknowledge CBP IT test platform (ENS de Lyon, France) for ML facilities and GPU devices, operating the SIDUS solution~\citep{SIDUS}. This work was granted access to the HPC resources of CINES under the DARI allocations A0050110575,  A0070110575, A0090110575 and A0110110575 made by GENCI.  We acknowledge the help of Alessandro Lovo in maintaining the GitHub page. {We would like to acknowledge the help of the referees in streamlining and improving the quality of this article.}

\section*{Competing interests}
Authors report no competing interests

\section{A list of abbreviations}

NWP - Numerical Weather Prediction
S2S - Subseasonal-to-Seasonal
SWG - Stochastic Weather Generator
GCM - General Circulation Model
CNN - Convolutional Neural Network
MCC - Matthew's Correlation Coefficient
VS - Validation Set
TS - Training Set
EVT - Extreme Value Theory
GEV - Generalized Extreme Value
NLS - Normalized Logarithmic Score
EOF - Empirical Orthogonal Function
PCA - Principal Component Analysis
VAE - Variational Autoencoder
CESM - Community Earth System Model
PlaSim - Planet Simulator

\bibliographystyle{apalike}
\bibliography{references.bib}

\begin{thebibliography}{}

\bibitem[Ailliot et~al., 2015]{ailliot2015stochastic}
Ailliot, P., Allard, D., Monbet, V., and Naveau, P. (2015).
\newblock Stochastic weather generators: an overview of weather type models.
\newblock {\em Journal de la Soci{\'e}t{\'e} Fran{\c{c}}aise de Statistique}, 156(1):101--113.

\bibitem[Balaji, 2021]{balaji}
Balaji, V. (2021).
\newblock Climbing down charney's ladder: machine learning and the post-dennard era of computational climate science.
\newblock {\em Phil. Trans.of the Royal Soc.A: Math., Phys.and Eng. Sciences}, 379(2194):20200085.

\bibitem[Barriopedro et~al., 2011]{barriopedro2011hot}
Barriopedro, D., Fischer, E.~M., Luterbacher, J., Trigo, R.~M., and Garc{\'\i}a-Herrera, R. (2011).
\newblock The hot summer of 2010: redrawing the temperature record map of europe.
\newblock {\em Science}, 332(6026):220--224.

\bibitem[Benedetti, 2010]{benedetti10}
Benedetti, R. (2010).
\newblock Scoring rules for forecast verification.
\newblock {\em Monthly Weather Review}, 138(1):203 -- 211.

\bibitem[Benson and Dirmeyer, 2021]{benson2021characterizing}
Benson, D.~O. and Dirmeyer, P.~A. (2021).
\newblock Characterizing the relationship between temperature and soil moisture extremes and their role in the exacerbation of heat waves over the contiguous united states.
\newblock {\em Journal of Climate}, 34(6):2175--2187.

\bibitem[Berg et~al., 2015]{Berg15}
Berg, A., Lintner, B.~R., Findell, K., Seneviratne, S.~I., van~den Hurk, B., Ducharne, A., Chéruy, F., Hagemann, S., Lawrence, D.~M., Malyshev, S., Meier, A., and Gentine, P. (2015).
\newblock Interannual coupling between summertime surface temperature and precipitation over land: Processes and implications for climate change.
\newblock {\em Journal of Climate}, 28(3):1308 -- 1328.

\bibitem[Besombes et~al., 2021]{besombes2021producing}
Besombes, C., Pannekoucke, O., Lapeyre, C., Sanderson, B., and Thual, O. (2021).
\newblock Producing realistic climate data with generative adversarial networks.
\newblock {\em Nonlinear Processes in Geophysics}, 28(3):347--370.

\bibitem[Beyer et~al., 1999]{beyer1999nearest}
Beyer, K., Goldstein, J., Ramakrishnan, R., and Shaft, U. (1999).
\newblock When is “nearest neighbor” meaningful?
\newblock In {\em International conference on database theory}, pages 217--235. Springer.

\bibitem[Bhatia et~al., 2021]{Bhatia21}
Bhatia, S., Jain, A., and Hooi, B. (2021).
\newblock Exgan: Adversarial generation of extreme samples.
\newblock {\em Proceedings of the AAAI Conference on Artificial Intelligence}, 35(8):6750--6758.

\bibitem[Bjerknes, 1921]{bjerknes1921meteorology}
Bjerknes, V. (1921).
\newblock The meteorology of the temperate zone and the general atmospheric circulation.
\newblock {\em Monthly weather review}, 49(1):1--3.

\bibitem[Boulaguiem et~al., 2022]{boulaguiem22}
Boulaguiem, Y., Zscheischler, J., Vignotto, E., van~der Wiel, K., and Engelke, S. (2022).
\newblock Modeling and simulating spatial extremes by combining extreme value theory with generative adversarial networks.
\newblock {\em Environmental Data Science}, 1:e5.

\bibitem[Buishand and Brandsma, 2001]{buishand2001multisite}
Buishand, T.~A. and Brandsma, T. (2001).
\newblock Multisite simulation of daily precipitation and temperature in the rhine basin by nearest-neighbor resampling.
\newblock {\em Water Resources Research}, 37(11):2761--2776.

\bibitem[Chattopadhyay et~al., 2020]{Chattopadhyay19}
Chattopadhyay, A., Nabizadeh, E., and Hassanzadeh, P. (2020).
\newblock Analog forecasting of extreme-causing weather patterns using deep learning.
\newblock {\em Journal of Advances in Modeling Earth Systems}, 12(2):e2019MS001958.
\newblock e2019MS001958 10.1029/2019MS001958.

\bibitem[Chen et~al., 2022]{Chen22}
Chen, K., Kuang, C., Wang, L., Chen, K., Han, X., and Fan, J. (2022).
\newblock Storm surge prediction based on long short-term memory neural network in the east china sea.
\newblock {\em Applied Sciences}, 12(1).

\bibitem[Choi et~al., 2021]{Choi2021}
Choi, W., Ho, C.-H., Jung, J., Chang, M., and Ha, K.-J. (2021).
\newblock Synoptic conditions controlling the seasonal onset and days of heatwaves over korea.
\newblock {\em Climate Dynamics}, 57(11):3045--3053.

\bibitem[Chraibi et~al., 2021]{Chraibi21}
Chraibi, H., Dutfoy, A., Galtier, T., and Garnier, J. (2021).
\newblock Optimal potential functions for the interacting particle system method.
\newblock {\em Monte Carlo Methods and Applications}, 27(2):137--152.

\bibitem[Christidis et~al., 2020]{Christidis2020}
Christidis, N., McCarthy, M., and Stott, P.~A. (2020).
\newblock The increasing likelihood of temperatures above 30 to 40 $\,^{\circ}$c in the united kingdom.
\newblock {\em Nature Communications}, 11(1):3093.

\bibitem[Cohen et~al., 2019]{cohen19}
Cohen, J., Coumou, D., Hwang, J., Mackey, L., Orenstein, P., Totz, S., and Tziperman, E. (2019).
\newblock S2s reboot: An argument for greater inclusion of machine learning in subseasonal to seasonal forecasts.
\newblock {\em WIREs Climate Change}, 10(2):e00567.

\bibitem[Correoso, 2019]{skextremes}
Correoso, K. (2019).
\newblock scikit-extremes.
\newblock \url{https://scikit-extremes.readthedocs.io/en/latest/}.

\bibitem[D'Andrea et~al., 2006]{dandrea06}
D'Andrea, F., Provenzale, A., Vautard, R., and De~Noblet-Decoudré, N. (2006).
\newblock Hot and cool summers: Multiple equilibria of the continental water cycle.
\newblock {\em Geophysical Research Letters}, 33(24).

\bibitem[Davis, 2002]{Davis02}
Davis, J.~C. (2002).
\newblock {\em Statistics and Data Analysis in Geology}.
\newblock Wiley, 3 edition.

\bibitem[Ding et~al., 2019]{ding2019diagnosing}
Ding, H., Newman, M., Alexander, M.~A., and Wittenberg, A.~T. (2019).
\newblock Diagnosing secular variations in retrospective enso seasonal forecast skill using cmip5 model-analogs.
\newblock {\em Geophysical Research Letters}, 46(3):1721--1730.

\bibitem[Doersch, 2016]{doersch2016tutorial}
Doersch, C. (2016).
\newblock Tutorial on variational autoencoders.
\newblock {\em arXiv preprint arXiv:1606.05908}.

\bibitem[E.~Quemener, 2014]{SIDUS}
E.~Quemener, M. (2014).
\newblock {"SIDUS", the solution for extreme deduplication of an operating system}.
\newblock {\em The Linux Journal}.

\bibitem[Felsche et~al., 2023]{felsche23}
Felsche, E., B{\"o}hnisch, A., and Ludwig, R. (2023).
\newblock Inter-seasonal connection of typical european heatwave patterns to soil moisture.
\newblock {\em npj Climate and Atmospheric Science}, 6(1):1.

\bibitem[Field et~al., 2012]{field2012managing}
Field, C.~B., Barros, V., Stocker, T.~F., and Dahe, Q. (2012).
\newblock {\em Managing the risks of extreme events and disasters to advance climate change adaptation: special report of the intergovernmental panel on climate change}.
\newblock Cambridge University Press.

\bibitem[Finkel et~al., 2023]{finkel23a}
Finkel, J., Gerber, E.~P., Abbot, D.~S., and Weare, J. (2023).
\newblock Revealing the statistics of extreme events hidden in short weather forecast data.
\newblock {\em AGU Advances}, 4(2):e2023AV000881.
\newblock e2023AV000881 2023AV000881.

\bibitem[Fischer et~al., 2007]{fischer2007soil}
Fischer, E.~M., Seneviratne, S.~I., Vidale, P.~L., L{\"u}thi, D., and Sch{\"a}r, C. (2007).
\newblock Soil moisture--atmosphere interactions during the 2003 european summer heat wave.
\newblock {\em Journal of Climate}, 20(20):5081--5099.

\bibitem[Fraedrich et~al., 2005]{fraedrich2005planet}
Fraedrich, K., Jansen, H., Kirk, E., Luksch, U., and Lunkeit, F. (2005).
\newblock The planet simulator: Towards a user friendly model.
\newblock {\em Meteorologische Zeitschrift}, 14(3):299--304.

\bibitem[Fraedrich et~al., 1998]{fraedrich_1998}
Fraedrich, K., Kirk, E., and Lunkeit, F. (1998).
\newblock Puma: Portable university model of the atmosphere.
\newblock {\em Deutsches Klimarechenzentrum}, page~38.

\bibitem[García-Herrera et~al., 2010]{Herrera2010}
García-Herrera, R., Díaz, J., Trigo, R.~M., Luterbacher, J., and Fischer, E.~M. (2010).
\newblock A review of the european summer heat wave of 2003.
\newblock {\em Critical Reviews in Environmental Science and Technology}, 40(4):267--306.

\bibitem[Gessner et~al., 2021]{gessner21}
Gessner, C., Fischer, E.~M., Beyerle, U., and Knutti, R. (2021).
\newblock Very rare heat extremes: Quantifying and understanding using ensemble reinitialization.
\newblock {\em Journal of Climate}, 34(16):6619 -- 6634.

\bibitem[Grönquist et~al., 2021]{gronquist_deep_2021}
Grönquist, P., Yao, C., Ben-Nun, T., Dryden, N., Dueben, P., Li, S., and Hoefler, T. (2021).
\newblock Deep learning for post-processing ensemble weather forecasts.
\newblock {\em Philosophical Transactions of the Royal Society A: Mathematical, Physical and Engineering Sciences}, 379(2194):20200092.

\bibitem[Ham et~al., 2019]{Ham}
Ham, Y.-G., Kim, J.-H., and Luo, J.-J. (2019).
\newblock Deep learning for multi-year enso forecasts.
\newblock {\em Nature}, 573(7775):568--572.

\bibitem[Hirschi et~al., 2011]{hirschi2011observational}
Hirschi, M., Seneviratne, S.~I., Alexandrov, V., Boberg, F., Boroneant, C., Christensen, O.~B., Formayer, H., Orlowsky, B., and Stepanek, P. (2011).
\newblock Observational evidence for soil-moisture impact on hot extremes in southeastern europe.
\newblock {\em Nature Geoscience}, 4(1):17--21.

\bibitem[Horowitz et~al., 2022]{horowitz2022circulation}
Horowitz, R.~L., McKinnon, K.~A., and Simpson, I.~R. (2022).
\newblock Circulation and soil moisture contributions to heatwaves in the united states.
\newblock {\em Journal of Climate}, 35(24):4431--4448.

\bibitem[Horton et~al., 2016]{horton2016review}
Horton, R.~M., Mankin, J.~S., Lesk, C., Coffel, E., and Raymond, C. (2016).
\newblock A review of recent advances in research on extreme heat events.
\newblock {\em Current Climate Change Reports}, 2(4):242--259.

\bibitem[Jacques-Dumas et~al., 2022]{jacques-dumas22}
Jacques-Dumas, V., Ragone, F., Borgnat, P., Abry, P., and Bouchet, F. (2022).
\newblock Deep learning-based extreme heatwave forecast.
\newblock {\em Frontiers in Climate}, 4.

\bibitem[Jacques-Dumas et~al., 2023]{jacques2022data}
Jacques-Dumas, V., van Westen, R.~M., Bouchet, F., and Dijkstra, H.~A. (2023).
\newblock Data-driven methods to estimate the committor function in conceptual ocean models.
\newblock {\em Nonlinear Processes in Geophysics}, 30(2):195--216.

\bibitem[J{\'e}z{\'e}quel et~al., 2018]{jezequel2018role}
J{\'e}z{\'e}quel, A., Yiou, P., and Radanovics, S. (2018).
\newblock Role of circulation in european heatwaves using flow analogues.
\newblock {\em Climate dynamics}, 50(3):1145--1159.

\bibitem[Karlsson and Yakowitz, 1987]{karlsson1987nearest}
Karlsson, M. and Yakowitz, S. (1987).
\newblock Nearest-neighbor methods for nonparametric rainfall-runoff forecasting.
\newblock {\em Water Resources Research}, 23(7):1300--1308.

\bibitem[Kingma and Welling, 2013]{kingma13}
Kingma, D.~P. and Welling, M. (2013).
\newblock Auto-encoding variational bayes.
\newblock {\em arXiv preprint arXiv:1312.6114}.

\bibitem[Koster and Suarez, 2001]{koster2001soil}
Koster, R.~D. and Suarez, M.~J. (2001).
\newblock Soil moisture memory in climate models.
\newblock {\em Journal of hydrometeorology}, 2(6):558--570.

\bibitem[Krouma et~al., 2023]{krouma23}
Krouma, M., Silini, R., and Yiou, P. (2023).
\newblock Ensemble forecast of an index of the madden--julian oscillation using a stochastic weather generator based on circulation analogs.
\newblock {\em Earth System Dynamics}, 14(1):273--290.

\bibitem[Lall and Sharma, 1996]{lall1996nearest}
Lall, U. and Sharma, A. (1996).
\newblock A nearest neighbor bootstrap for resampling hydrologic time series.
\newblock {\em Water resources research}, 32(3):679--693.

\bibitem[Lestang et~al., 2018]{Lestang_2018}
Lestang, T., Ragone, F., Br{\'{e}}hier, C.-E., Herbert, C., and Bouchet, F. (2018).
\newblock Computing return times or return periods with rare event algorithms.
\newblock {\em Journal of Statistical Mechanics: Theory and Experiment}, 2018(4):043213.

\bibitem[Lguensat et~al., 2017]{lguensat2017}
Lguensat, R., Tandeo, P., Ailliot, P., Pulido, M., and Fablet, R. (2017).
\newblock The analog data assimilation.
\newblock {\em Monthly Weather Review}, 145(10):4093 -- 4107.

\bibitem[Loaiza-Ganem and Cunningham, 2019]{loaiza2019continuous}
Loaiza-Ganem, G. and Cunningham, J.~P. (2019).
\newblock The continuous bernoulli: fixing a pervasive error in variational autoencoders.
\newblock {\em Advances in Neural Information Processing Systems}, 32.

\bibitem[Lopez-Gomez et~al., 2022]{lopez2022global}
Lopez-Gomez, I., McGovern, A., Agrawal, S., and Hickey, J. (2022).
\newblock Global extreme heat forecasting using neural weather models.
\newblock {\em arXiv preprint arXiv:2205.10972}.

\bibitem[Lorenz, 1969]{lorenz69}
Lorenz, E.~N. (1969).
\newblock Atmospheric predictability as revealed by naturally occurring analogues.
\newblock {\em Journal of Atmospheric Sciences}, 26(4):636 -- 646.

\bibitem[Lorenz et~al., 2010]{lorenz2010persistence}
Lorenz, R., Jaeger, E.~B., and Seneviratne, S.~I. (2010).
\newblock Persistence of heat waves and its link to soil moisture memory.
\newblock {\em Geophysical Research Letters}, 37(9).

\bibitem[Lucente et~al., 2022a]{Lucente2022committor}
Lucente, D., Herbert, C., and Bouchet, F. (2022a).
\newblock Committor functions for climate phenomena at the predictability margin: The example of el niño southern oscillation in the jin and timmermann model.
\newblock {\em Journal of the Atmospheric Sciences}.

\bibitem[Lucente et~al., 2022b]{lucente2022coupling}
Lucente, D., Rolland, J., Herbert, C., and Bouchet, F. (2022b).
\newblock Coupling rare event algorithms with data-based learned committor functions using the analogue markov chain.
\newblock {\em Journal of Statistical Mechanics: Theory and Experiment}, 2022(8):083201.

\bibitem[Manabe, 1969]{manabe69}
Manabe, S. (1969).
\newblock Climate and the ocean circulation: I. the atmosphere circulation and the hydrology of the earth's surface.
\newblock {\em Monthly Weather Review}, 97(11):739 -- 774.

\bibitem[Matthews, 1975]{MATTHEWS1975}
Matthews, B. (1975).
\newblock Comparison of the predicted and observed secondary structure of t4 phage lysozyme.
\newblock {\em Biochimica et Biophysica Acta (BBA) - Protein Structure}, 405(2):442--451.

\bibitem[Mehta et~al., 2019]{MEHTA20191}
Mehta, P., Bukov, M., Wang, C.-H., Day, A.~G., Richardson, C., Fisher, C.~K., and Schwab, D.~J. (2019).
\newblock A high-bias, low-variance introduction to machine learning for physicists.
\newblock {\em Physics Reports}, 810:1--124.
\newblock A high-bias, low-variance introduction to Machine Learning for physicists.

\bibitem[Miloshevich et~al., 2023a]{miloshevich22}
Miloshevich, G., Cozian, B., Abry, P., Borgnat, P., and Bouchet, F. (2023a).
\newblock Probabilistic forecasts of extreme heatwaves using convolutional neural networks in a regime of lack of data.
\newblock {\em Phys. Rev. Fluids}, 8:040501.

\bibitem[Miloshevich et~al., 2023b]{miloshevich2023robust}
Miloshevich, G., Rouby-Poizat, P., Ragone, F., and Bouchet, F. (2023b).
\newblock Robust intra-model teleconnection patterns for extreme heatwaves.
\newblock {\em Frontiers in Earth Science}, 11.

\bibitem[Min et~al., 2020]{ki20}
Min, S.-K., Kim, Y.-H., Lee, S.-M., Sparrow, S., Li, S., Lott, F.~C., and Stott, P.~A. (2020).
\newblock Quantifying human impact on the 2018 summer longest heat wave in south korea.
\newblock {\em Bulletin of the American Meteorological Society}, 101(1):S103 -- S108.

\bibitem[Miralles et~al., 2019]{miralles19}
Miralles, D.~G., Gentine, P., Seneviratne, S.~I., and Teuling, A.~J. (2019).
\newblock Land–atmospheric feedbacks during droughts and heatwaves: state of the science and current challenges.
\newblock {\em Annals of the New York Academy of Sciences}, 1436(1):19--35.

\bibitem[Miralles et~al., 2014]{miralles14}
Miralles, D.~G., Teuling, A.~J., van Heerwaarden, C.~C., and Vil{\`a}-Guerau~de Arellano, J. (2014).
\newblock Mega-heatwave temperatures due to combined soil desiccation and atmospheric heat accumulation.
\newblock {\em Nature Geoscience}, 7(5):345--349.

\bibitem[{National Academies of Sciences Engineering and Medicine}, 2016]{national_academies_of_sciences_engineering_and_medicine_attribution_2016}
{National Academies of Sciences Engineering and Medicine}, editor (2016).
\newblock {\em Attribution of {Extreme} {Weather} {Events} in the {Context} of {Climate} {Change}}.
\newblock The National Academies Press, Washington, DC.

\bibitem[Ragone and Bouchet, 2021]{Ragone21}
Ragone, F. and Bouchet, F. (2021).
\newblock Rare event algorithm study of extreme warm summers and heatwaves over europe.
\newblock {\em Geophysical Research Letters}, 48(12):e2020GL091197.
\newblock e2020GL091197 2020GL091197.

\bibitem[Ragone et~al., 2018]{Ragone18}
Ragone, F., Wouters, J., and Bouchet, F. (2018).
\newblock Computation of extreme heat waves in climate models using a large deviation algorithm.
\newblock {\em Proceedings of the National Academy of Sciences}, 115(1):24--29.

\bibitem[Rajagopalan and Lall, 1999]{rajagopalan1999k}
Rajagopalan, B. and Lall, U. (1999).
\newblock A k-nearest-neighbor simulator for daily precipitation and other weather variables.
\newblock {\em Water resources research}, 35(10):3089--3101.

\bibitem[Reichstein et~al., 2019]{reichstein19}
Reichstein, M., Camps-Valls, G., Stevens, B., Jung, M., Denzler, J., Carvalhais, N., and Prabhat (2019).
\newblock Deep learning and process understanding for data-driven earth system science.
\newblock {\em Nature}, 566(7743):195--204.

\bibitem[Rezende et~al., 2014]{rezende14}
Rezende, D.~J., Mohamed, S., and Wierstra, D. (2014).
\newblock Stochastic backpropagation and approximate inference in deep generative models.
\newblock In Xing, E.~P. and Jebara, T., editors, {\em Proceedings of the 31st International Conference on Machine Learning}, volume~32 of {\em Proceedings of Machine Learning Research}, pages 1278--1286, Bejing, China. PMLR.

\bibitem[Rowntree and Bolton, 1983]{rowntree1983simulation}
Rowntree, P. and Bolton, J. (1983).
\newblock Simulation of the atmospheric response to soil moisture anomalies over europe.
\newblock {\em Quarterly Journal of the Royal Meteorological Society}, 109(461):501--526.

\bibitem[Russo et~al., 2015]{russo2015top}
Russo, S., Sillmann, J., and Fischer, E.~M. (2015).
\newblock Top ten european heatwaves since 1950 and their occurrence in the coming decades.
\newblock {\em Environmental Research Letters}, 10(12):124003.

\bibitem[Schubert et~al., 2014]{schubert2014northern}
Schubert, S.~D., Wang, H., Koster, R.~D., Suarez, M.~J., and Groisman, P.~Y. (2014).
\newblock Northern eurasian heat waves and droughts.
\newblock {\em Journal of Climate}, 27(9):3169--3207.

\bibitem[Schulz and Lerch, 2022]{shulz22}
Schulz, B. and Lerch, S. (2022).
\newblock Machine learning methods for postprocessing ensemble forecasts of wind gusts: A systematic comparison.
\newblock {\em Monthly Weather Review}, 150(1):235 -- 257.

\bibitem[Seneviratne et~al., 2012]{Seneviratne12}
Seneviratne, S., Nicholls, N., Easterling, D., Goodess, C., Kanae, S., Kossin, J., Luo, Y., Marengo, J., McInnes, K., Rahimi, M., Reichstein, M., Sorteberg, A., Vera, C., and Zhang, X. (2012).
\newblock Changes in climate extremes and their impacts on the natural physical environment. in: Managing the risks of extreme events and disasters to advance climate change adaptation.
\newblock {\em A Special Report of Working Groups I and II of the IPCC}, pages 109--230.

\bibitem[Seneviratne et~al., 2021]{IPCC_2021_extremes}
Seneviratne, S., Zhang, X., Adnan, M., Badi, W., Dereczynski, C., Di~Luca, A., Ghosh, S., Iskandar, I., Kossin, J., Lewis, S., Otto, F., Pinto, I., Satoh, M., Vicente-Serrano, S., Wehner, M., and Zhou, B. (2021).
\newblock {\em Weather and Climate Extreme Events in a Changing Climate}, page 1513–1766.
\newblock Cambridge University Press, Cambridge, United Kingdom and New York, NY, USA.

\bibitem[Seneviratne et~al., 2010]{seneviratne2010investigating}
Seneviratne, S.~I., Corti, T., Davin, E.~L., Hirschi, M., Jaeger, E.~B., Lehner, I., Orlowsky, B., and Teuling, A.~J. (2010).
\newblock Investigating soil moisture--climate interactions in a changing climate: A review.
\newblock {\em Earth-Science Reviews}, 99(3-4):125--161.

\bibitem[Seneviratne et~al., 2006]{seneviratne2006soil}
Seneviratne, S.~I., Koster, R.~D., Guo, Z., Dirmeyer, P.~A., Kowalczyk, E., Lawrence, D., Liu, P., Mocko, D., Lu, C.-H., Oleson, K.~W., et~al. (2006).
\newblock Soil moisture memory in agcm simulations: analysis of global land--atmosphere coupling experiment (glace) data.
\newblock {\em Journal of Hydrometeorology}, 7(5):1090--1112.

\bibitem[Shukla and Mintz, 1982]{shukla1982influence}
Shukla, J. and Mintz, Y. (1982).
\newblock Influence of land-surface evapotranspiration on the earth's climate.
\newblock {\em Science}, 215(4539):1498--1501.

\bibitem[Sohn et~al., 2015]{sohn15}
Sohn, K., Lee, H., and Yan, X. (2015).
\newblock Learning structured output representation using deep conditional generative models.
\newblock In Cortes, C., Lawrence, N., Lee, D., Sugiyama, M., and Garnett, R., editors, {\em Advances in Neural Information Processing Systems}, volume~28. Curran Associates, Inc.

\bibitem[St\'{e}fanon et~al., 2012]{Stefanon_2012}
St\'{e}fanon, M., D'Andrea, F., and Drobinski, P. (2012).
\newblock Heatwave classification over europe and the mediterranean region.
\newblock {\em Environmental Research Letters}, 7:014023.

\bibitem[Stephenson, 1997]{stephenson1997correlation}
Stephenson, D. (1997).
\newblock Correlation of spatial climate/weather maps and the advantages of using the mahalanobis metric in predictions.
\newblock {\em Tellus A}, 49(5):513--527.

\bibitem[Thuerey et~al., 2021]{thuerey2021pbdl}
Thuerey, N., Holl, P., Mueller, M., Schnell, P., Trost, F., and Um, K. (2021).
\newblock {\em Physics-based Deep Learning}.
\newblock WWW.

\bibitem[Thuerey et~al., 2020]{thuerey20}
Thuerey, N., Wei\ss{}enow, K., Prantl, L., and Hu, X. (2020).
\newblock Deep learning methods for reynolds-averaged navier–stokes simulations of airfoil flows.
\newblock {\em AIAA Journal}, 58(1):25--36.

\bibitem[van~den Dool, 2007]{vandenDool07}
van~den Dool, H. (2007).
\newblock {\em Empirical Methods in Short-Term Climate Prediction}.
\newblock Oxford University Press, USA.

\bibitem[Van~den Dool et~al., 2003]{van2003performance}
Van~den Dool, H., Huang, J., and Fan, Y. (2003).
\newblock Performance and analysis of the constructed analogue method applied to us soil moisture over 1981--2001.
\newblock {\em Journal of Geophysical Research: Atmospheres}, 108(D16).

\bibitem[van Straaten et~al., 2022]{straaten22}
van Straaten, C., Whan, K., Coumou, D., van~den Hurk, B., and Schmeits, M. (2022).
\newblock Using explainable machine learning forecasts to discover subseasonal drivers of high summer temperatures in western and central europe.
\newblock {\em Monthly Weather Review}, 150(5):1115 -- 1134.

\bibitem[Vargas~Zeppetello and Battisti, 2020]{vargas2020projected}
Vargas~Zeppetello, L. and Battisti, D. (2020).
\newblock Projected increases in monthly midlatitude summertime temperature variance over land are driven by local thermodynamics.
\newblock {\em Geophysical Research Letters}, 47(19):e2020GL090197.

\bibitem[Vautard et~al., 2007]{vautard2007}
Vautard, R., Yiou, P., D'Andrea, F., {de Noblet}, N., Viovy, N., Cassou, C., Polcher, J., Ciais, P., Kageyama, M., and Fan, Y. (2007).
\newblock Summertime {{European}} heat and drought waves induced by wintertime {{Mediterranean}} rainfall deficit.
\newblock {\em Geophys. Res. Lett.}, 34(7):L07711.

\bibitem[Wang et~al., 2020]{wang2020extended}
Wang, X., Slawinska, J., and Giannakis, D. (2020).
\newblock Extended-range statistical enso prediction through operator-theoretic techniques for nonlinear dynamics.
\newblock {\em Scientific reports}, 10(1):1--15.

\bibitem[Watson, 2022]{watson_machine_2022}
Watson, P. A.~G. (2022).
\newblock Machine learning applications for weather and climate need greater focus on extremes.
\newblock {\em Environmental Research Letters}, 17(11):111004.
\newblock Publisher: {IOP} Publishing.

\bibitem[Webber et~al., 2019]{webber19}
Webber, R.~J., Plotkin, D.~A., O’Neill, M.~E., Abbot, D.~S., and Weare, J. (2019).
\newblock Practical rare event sampling for extreme mesoscale weather.
\newblock {\em Chaos: An Interdisciplinary Journal of Nonlinear Science}, 29(5):053109.

\bibitem[Wilks, 1992]{wilks1992adapting}
Wilks, D.~S. (1992).
\newblock Adapting stochastic weather generation algorithms for climate change studies.
\newblock {\em Climatic change}, 22(1):67--84.

\bibitem[Wilks, 2019]{wilks19}
Wilks, D.~S., editor (2019).
\newblock {\em Statistical Methods in the Atmospheric Sciences (Fourth Edition)}.
\newblock Elsevier, fourth edition edition.

\bibitem[Yates et~al., 2003]{yates2003technique}
Yates, D., Gangopadhyay, S., Rajagopalan, B., and Strzepek, K. (2003).
\newblock A technique for generating regional climate scenarios using a nearest-neighbor algorithm.
\newblock {\em Water Resources Research}, 39(7).

\bibitem[Yiou, 2014]{yiou2014anawege}
Yiou, P. (2014).
\newblock Anawege: a weather generator based on analogues of atmospheric circulation.
\newblock {\em Geoscientific Model Development}, 7(2):531--543.

\bibitem[Yiou et~al., 2023]{yiou:hal-03921111}
Yiou, P., Cadiou, C., Faranda, D., J{\'e}z{\'e}quel, A., Malhomme, N., Miloshevich, G., Noyelle, R., Pons, F., Robin, Y., and Vrac, M. (2023).
\newblock {Worst case climate simulations show that heatwaves could cause major disruptions in the Paris 2024 Olympics}.
\newblock working paper or preprint.

\bibitem[Zeppetello et~al., 2022]{zeppetello2022physics}
Zeppetello, L. R.~V., Battisti, D.~S., and Baker, M.~B. (2022).
\newblock The physics of heat waves: What causes extremely high summertime temperatures?
\newblock {\em Journal of Climate}, 35(7):2231--2251.

\bibitem[Zhang and Boos, 2023]{zhang23}
Zhang, Y. and Boos, W.~R. (2023).
\newblock An upper bound for extreme temperatures over midlatitude land.
\newblock {\em Proceedings of the National Academy of Sciences}, 120(12):e2215278120.

\bibitem[Zhou et~al., 2019]{zhou2019land}
Zhou, S., Williams, A.~P., Berg, A.~M., Cook, B.~I., Zhang, Y., Hagemann, S., Lorenz, R., Seneviratne, S.~I., and Gentine, P. (2019).
\newblock Land--atmosphere feedbacks exacerbate concurrent soil drought and atmospheric aridity.
\newblock {\em Proceedings of the National Academy of Sciences}, 116(38):18848--18853.

\end{thebibliography}

\appendix
\section{Principal Component Analysis and Variational Autoencoder}\label{sec:vae_methods}

In Plasim we encounter high dimensional fields, compared to what is done in toy models such as Lorenz 63. Generally speaking one has to deal with problems of ``curse of dimensionality'' in such cases and the need to somehow project the dynamics $x\rightarrow z$.
{To be more specific, let $\mathcal{D}\in\mathbb{R}^{n,d}$ be a dataset consisting of $n$ samples $x\in\mathbb{R}^{d}$ and let $z=f(x)\in\mathbb{R}^p$ a lower dimensional projection of the data ($p\ll d$). In what follows we will give two brief descriptions of two widely used projection techniques involving the use of both linear and non-linear functions $f$, namely  Principal Component Analysis (PCA) and Variational Autoencoder (VAE). PCA (also known as Empirical Orthogonal Functions (EOFs)) consists of a linear transformation of the data points $x$ which aims at preserving as much variance as possible of the original dataset. To do so,  $x$ is projected onto the low-dimensional space spanned by the first $p$-eigenvectors of the correlation matrix $\Sigma=\mathcal{D}^{T}\mathcal{D}$\footnote{The first $p$-eigenvectors are those corresponding to the largest $p$ eigenvalues. }. The dimensionality $p$ of the latent space is determined by the percentage of data variability that is desired to be preserved by the transformation. Here however dimensionality reduction is used as a pre-processing step in order to optimize a prediction task. From this perspective, not all information regarding the variability of observations is necessarily relevant to the forecasting task.}

{The second projection technique we want to discuss consists of a nonlinear dimensionality reduction via variational autoencoder (VAE)~\citep{kingma13}.}
The idea is to project fields such as geopotential $x$ on the latent low dimensional space $z$, from where realistic looking states can be sampled using, for instance, a Gaussian measure. VAE consists of probabilistic \emph{decoder} $p_\theta(x|z)$ and probabilistic (stochastic) \emph{encoder} $q_\phi(z|x)$. The goal is to maximize the probability of the data $p(x)$ in the training set~\citep{doersch2016tutorial}
\begin{equation}
    p(x) = \int dz\, p_\theta(x|z) p(z),
\end{equation}
where traditionally the prior $p(z) = \mathcal{N}(0,I)$. The choice for the likelihood given by the decoder is often chosen to be Gaussian $p(x|z) = \mathcal{N}\left(x|f_\theta(z),\sigma^2 \, I\right)$ or Bernoulli (which is particularly suitable when the underlying data is binary~\citep{loaiza2019continuous}) and $f_\theta(z)$ can be some neural network with weights $\theta$. Formally, posterior $p_\theta(z|x)$ could be written using Bayes's rule but in most practical cases this leads to intractable expression and methods such as importance sampling may lead to the variance in the estimate that can be very high if the proposal distribution is poor so instead Evidence Lowest Bound (ELBO) is used ~\citep{MEHTA20191}.  This leads to approximate posterior which is also parametrized using neural networks $q_\phi(z|x) = \mathcal{N}\left(z|\mu_\phi(x), \Sigma_\phi(x)\right)$. For details on how this is achieved, such as ``reparametrization trick'' see for example~\citep{rezende14}. Other variants of VAE have been developed such as conditional VAE~\citep{sohn15} but since we are looking for a space on which distance can be computed to allow simulating unconditional trajectories we believe the latent space should be also unconditional. VAEs are not the only probabilistic generative models, and indeed Generative Adverserial Networks (GANs) are often preferred choice since VAEs often produce smoothened images. On the other hand GANs may suffer from technical difficulties such as mode collapse and are generally more difficult to train.

Once the training is performed (on the training set) we may project the fields in both training and validation set as $\mathcal{Z}(\mathbf{r},t)\rightarrow \mathcal{V}_{\text{tr}}\, \overline{\mathcal{Z}}(\mathbf{r},t)$ to obtain latent variables ${\bf z}(t):=\{z_m(t),m\in 1,\dots, M\}\sim \mathcal{N}(0,1)$ that are stacked along with the area integrals
\begin{equation}\label{eq:vae_ZG}
    \mathcal{X} = \left({\bf z}(t), \langle\overline{\mathcal{T}}\rangle_{\mathcal{D}}(t), \langle\overline{\mathcal{S}}\rangle_{\mathcal{D}}(t)  \right),
\end{equation}

\begin{figure}
    \centering
    \includegraphics[width=\linewidth]{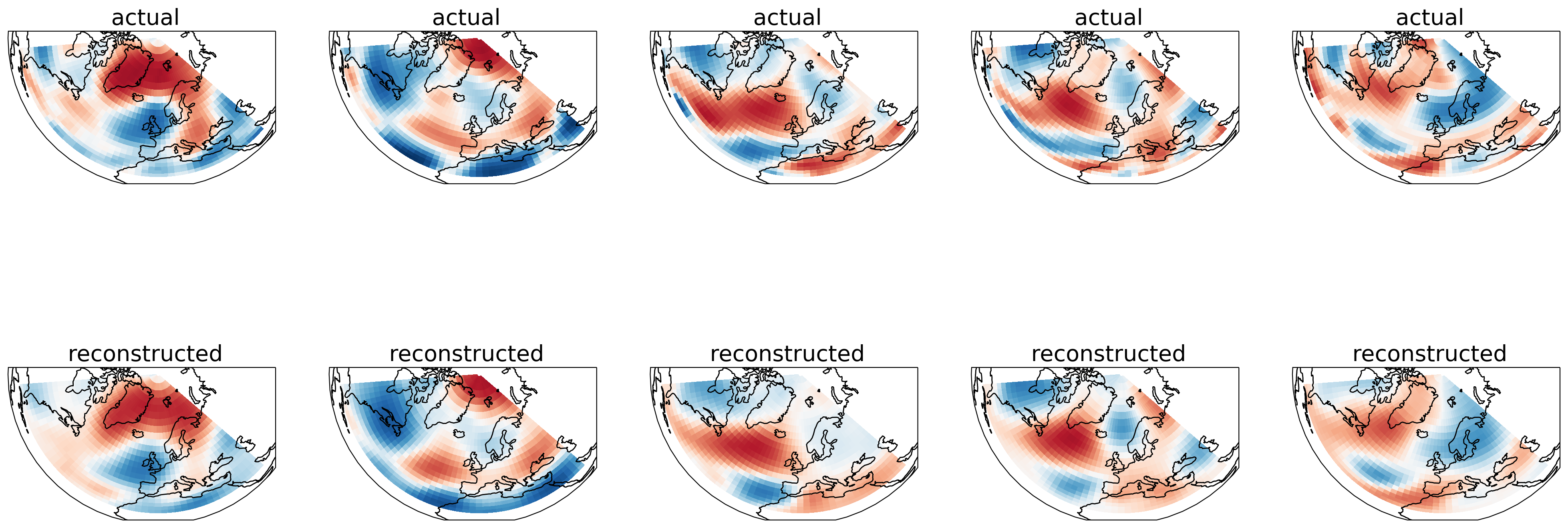}
    \caption{Top panels: original geopotential anomalies, Bottom panels: a realisation of reconstructed geopotential anomalies after passing them through VAE. }
    \label{fig:reconstruction}
\end{figure}

For this aim we use 8 layer encoder-decoder type network with residual connections within encoder and decoder respectively, which allows projecting $\mathcal{Z}$ to the 16 dimensional latent space. In this case, the input corresponding to $\mathcal{Z}$ field contribution in~Eq. \eqref{eq:metric} is provided from the latent space. The VAE is trained only on the Training Set (TS). The training takes approximately 50 epochs for each fold on the corresponding training set using TU102 [RTX 2080 Ti Rev. A] GPU which results in total training time of approximately one hour. Such large degree of freedom reduction naturally leads to significant gains (2 orders of magnitude) in subsequent kDTree computation speed. The forecast skill of this VAE SWG, however, remains the same compared to regular SWG. The original and VAE reconstructed geopotential anomalies are plotted for few samples weather situations on Figure~\ref{fig:reconstruction}.

\section{Daily steps}\label{sec:daily}

\begin{figure}
    \centering
    \includegraphics[width=.5\linewidth]{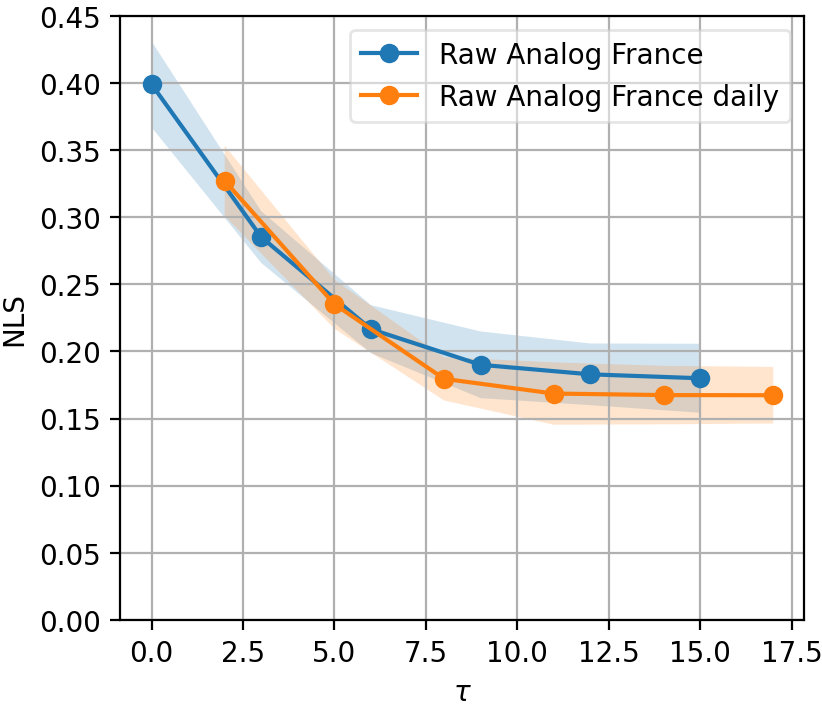}
    \caption{{\bf Basic SWG daily increments vs 3 day} Here we show NLS (equation~\eqref{eq:NLS}) as a function of lead time $\tau$ for optimal hyper-parameters obtained for usual analog Markov chain based on representation (equation~\eqref{eq:simple_markov}) and the procedure described in~\ref{sec:SWG_algorithm} like in Figure~\eqref{eq:NLS}. Both Figures share identical blue curves (corresponding to the same setup. Orange curve corresponds to a Markov chain whose increment $\tau_m$ and coarse-graining time $\tau_c$ are set to $1$, i.e. no coarse-graining. Note that for fairness of comparison we have shifted the orange and green curves by 2 days (see discussion in the Section~\ref{sec:NLS}).For details on the interpretation of different panels see caption of Figure~\ref{fig:tau_simple_analog}. }
    \label{fig:tau_simple_analog_percent}
\end{figure}

Throughout this paper we have used $\tau_m=\tau_c=3$ day step for our Markov chain and coarse graining. While the motivation for this was synoptic decorrelation time one naturally wonders how optimal is that choice for predicting extremes. Moreover, the CNN was designed to learn from daily fields, which meant we had to shift the NLS curves by $2$ days for proper comparisons. Here we will show what happens when $\tau_m=\tau_c=1$ day. Given that temperature autocorrelation time one would expect subsequent days to be strongly correlated, which would violate Markov's condition. 
Nevertheless, comparing blue and orange curves on Figure~\ref{fig:tau_simple_analog_percent} we see hardly any difference. 

\section{Committors for 100 years}\label{sec:committorD100}

\begin{figure}
    \centering
    \includegraphics[width=\linewidth]{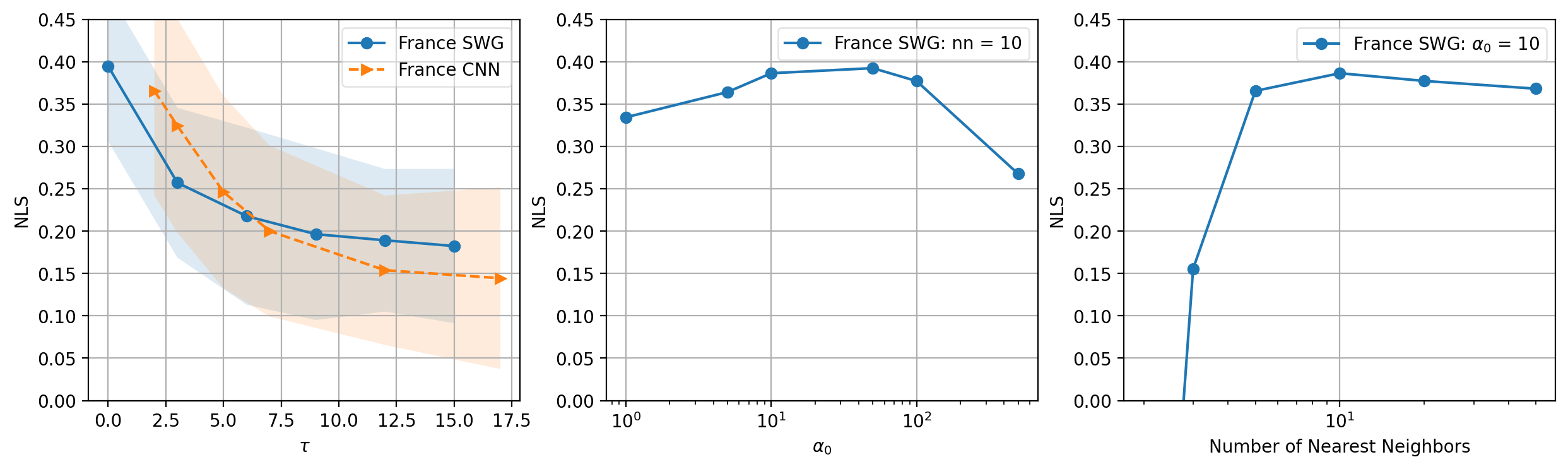}
    \includegraphics[width=\linewidth]{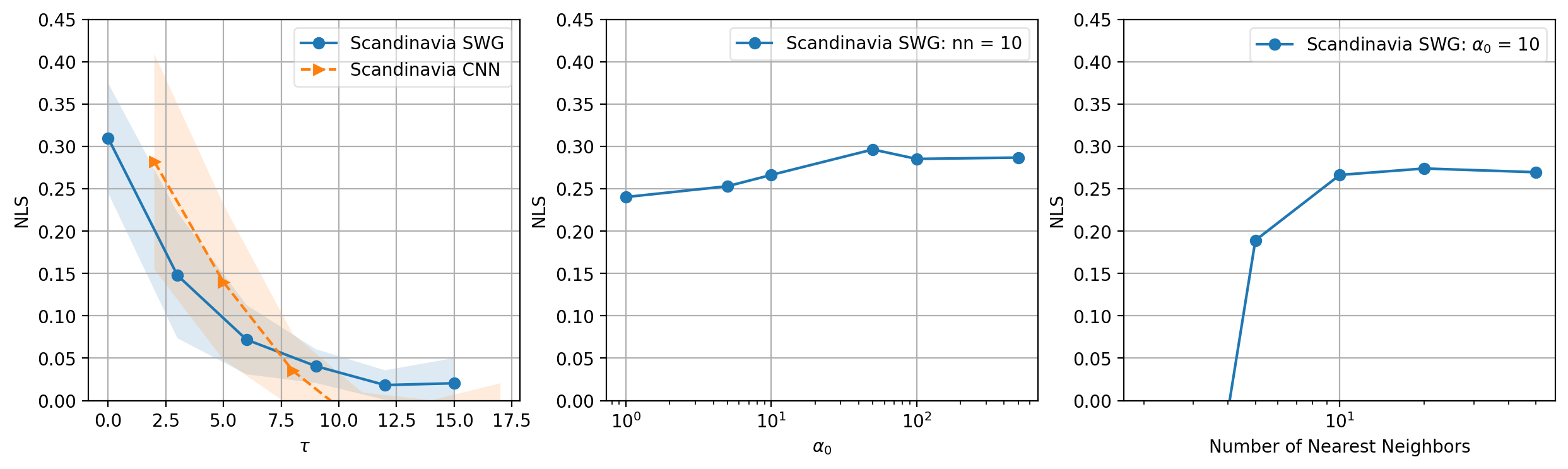}
    \caption{{\bf Basic SWG (blue curve) vs CNN. (orange curve)} This figure is identical to the Figure~\ref{fig:tau_simple_analog} except for the number of years used to train/validate the algorithm. We have relied on D100 here (Table~\ref{tab:datasets}).}
    \label{fig:tau_simple_analog_D100}
\end{figure}

Results of the Figure~\ref{fig:tau_simple_analog} show that CNN outperforms SWG when using D500 (see Table~\ref{tab:datasets}), in other words 400 years of training data. Climate models allow working with large datasets, however in practice we are often limited by the paucity of the observational record. Thus we provide benchmarks for shorter dataset D100, thus 80 years of training, the same that is used in Section~\ref{sec:sampling}. The results are plotted on Figure~\ref{fig:tau_simple_analog_D100} which shows that, while CNN still gives better results on average, the spread of the NLS tends to be quite large. The trends for $\alpha_0$ and number of neighbors remain the same. 

\section{Return times for different values of the hyperparameter}\label{sec:returnsalpha50}

\begin{figure}
    \centering
    \includegraphics[width=.9\textwidth]{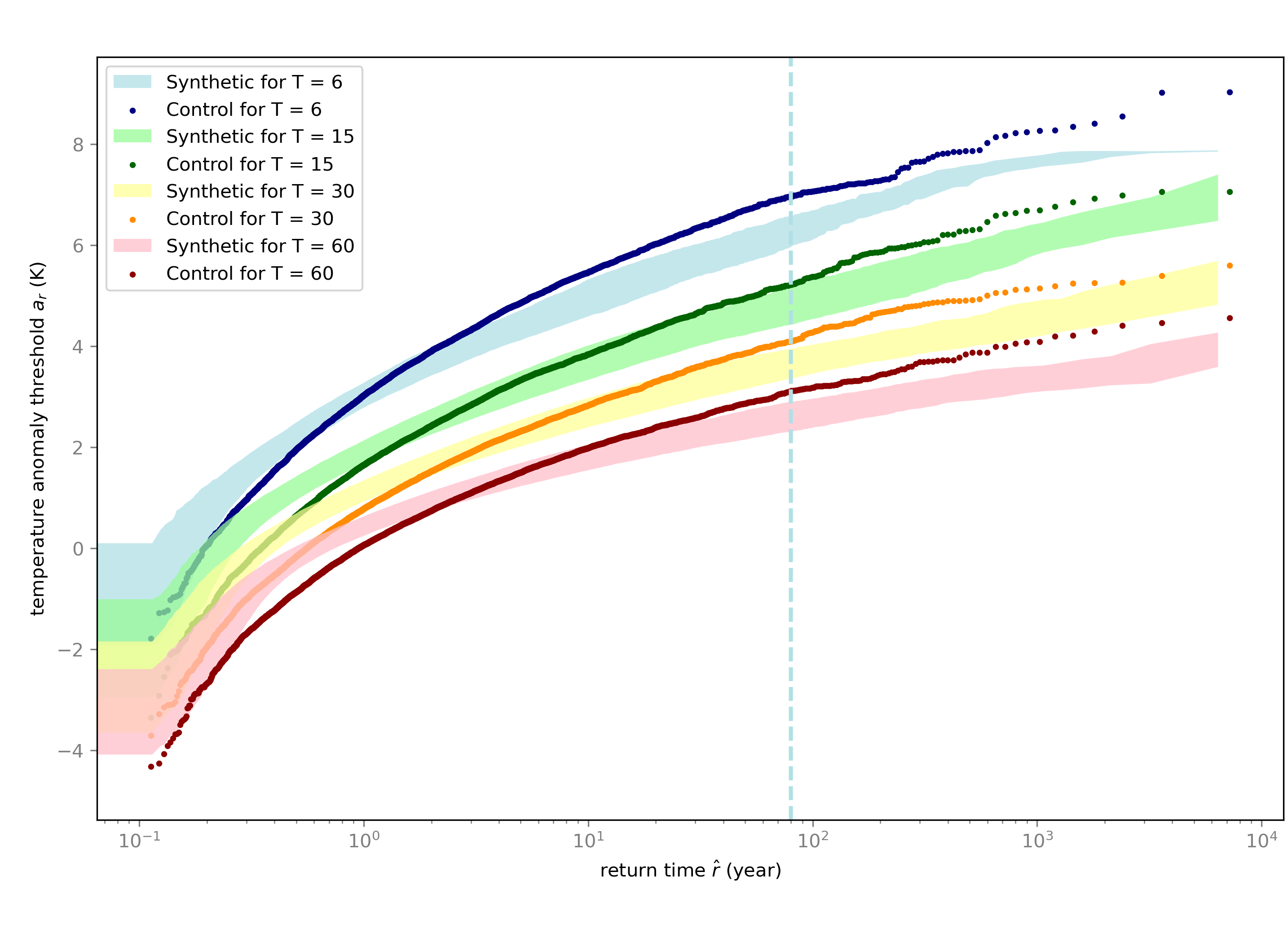}
    \caption{Return time plot for France heatwaves using analogues of North Atlantic and Europe. Here we use parameters $\alpha=50$ (default), Number of nearest neighbors $n=10$, the analogs are initialized on June 1 of each year (using the simulation data) and then advanced according to the Algorithm~\ref{alg:SWG}. The trajectory ends at the last day of summer. Each trajectory is sampled 800 times providing much longer synthetic series and thus estimating longer return times. Return times are computed for $T=\{6,15,30,60\}$ day heatwaves (indicated on the inset legend), with dots corresponding to the statistics from the control run (D8000, see Table~\ref{tab:datasets}), while shaded areas correspond to the bootstrapped synthetic trajectory: the whole sequence is split into 10 portions which allows estimating the mean and variance. The shading corresponds to mean plus or minus one standard deviation.  }
    \label{fig:FranceReturnsAlpha50}
\end{figure}

Here we attach Figure~\ref{fig:FranceReturnsAlpha50} which is equivalent to Figure~\ref{fig:FranceReturns} in all but the value of $\alpha_0=50$. This parameter was chosen optimal for this area in Section~\ref{sec:forecasting}. As we can see it is not optimal for generating synthetic return times (under-estimation). This under-estimation is related to the reduction of variance problem. We do not include the case of Scandinavia, but the results look very similar. 

\end{document}